\documentclass[twocolumn]{aastex63}

\newcommand{\kepler}{\textit{Kepler }}
\newcommand{\TESS}{\textit{TESS }}
\newcommand{\HIP}{\textit{Hipparcos }}
\newcommand{\Gaia}{\textit{Gaia }}
\newcommand{\masyr}{mas yr$^{-1}$}
\usepackage{CJK}
\usepackage{amsmath}
\usepackage{natbib}
\usepackage{graphicx}
\usepackage{amsmath}
\usepackage{booktabs}
\usepackage{longtable}
\usepackage{xspace}
\usepackage{hyperref}
\usepackage[T1]{fontenc}
\usepackage{lipsum}
\usepackage{comment}
\usepackage{xcolor}
\usepackage{chngcntr}

\received{June 1, 2022}
\revised{June 1, 2022}
\accepted{\today}

\shorttitle{S-Type Planets in Binaries}
\shortauthors{Zhang et al.}

\graphicspath{{./}{figures/}}

\begin{document}
\begin{CJK*}{UTF8}{gbsn}
\title{Dynamical Architectures of S-type Transiting Planets in Binaries I: Target Selection using Hipparcos and Gaia proper motion anomalies \footnote{Also referred to as Hipparcos and Gaia astrometric accelerations}}

\correspondingauthor{Jingwen Zhang}
\email{jingwen7@hawaii.edu}


\author[0000-0002-2696-2406]{Jingwen Zhang (张婧雯)}
\altaffiliation{NASA FINESST Fellow}
\affiliation{Institute for Astronomy, University of Hawai`i, Honolulu, HI 96822, USA}

\author[0000-0002-3725-3058]{Lauren M. Weiss}
\affiliation{Department of Physics, University of Notre Dame\\
334 Nieuwland Science Hall, Notre Dame, IN 46556, USA}

\author[0000-0001-8832-4488]{Daniel Huber}
\affiliation{Institute for Astronomy, University of Hawai`i, Honolulu, HI 96822, USA}

\author[0000-0002-4625-7333]{Eric L.\ N.\ Jensen}
\affiliation{Department of Physics \& Astronomy, Swarthmore College, Swarthmore PA 19081, USA}

\author[0000-0003-2630-8073]{Timothy D. Brandt}
\affiliation{Department of Physics, University of California, Santa Barbara, Santa Barbara, CA 93106, USA}

\author[0000-0001-6588-9574]{Karen Collins}
\affiliation{Center for Astrophysics \textbar \ Harvard \& Smithsonian, 60 Garden Street, Cambridge, MA 02138, USA}

\author[0000-0003-2239-0567]{Dennis M.\ Conti}
\affiliation{American Association of Variable Star Observers, 185 Alewife Brook Parkway, Suite 410, Cambridge, MA 02138, USA}

\author[0000-0002-0531-1073]{Howard Isaacson}
\affiliation{501 Campbell Hall, University of California at Berkeley, Berkeley, CA 94720, USA}
\affiliation{Centre for Astrophysics, University of Southern Queensland, Toowoomba, QLD, Australia}

\author[0000-0003-0828-6368]{Pablo Lewin}
\affiliation{The Maury Lewin Astronomical Observatory, Glendora,California.91741. USA}

\author[0000-0001-8134-0389]{Guiseppi Marino}
\affiliation{Wild Boar Remote Observatory, San Casciano in val di Pesa, Firenze, 50026 Italy}
\affiliation{INAF - Osservatorio Astrofisico di Catania, Via S. Sofia 78, 95123 Catania, Italy}

\author[0000-0001-8879-7138]{Bob Massey}
\affil{Villa '39 Observatory, Landers, CA 92285, USA}

\author[0000-0001-9087-1245]{Felipe Murgas}
\affiliation{Instituto de Astrof\'isica de Canarias (IAC), E-38205 La Laguna, Tenerife, Spain}
\affiliation{Departamento de Astrof\'isica, Universidad de La Laguna (ULL), E-38206 La Laguna, Tenerife, Spain}

\author[0000-0003-0987-1593]{Enric Palle}
\affiliation{Instituto de Astrof\'isica de Canarias (IAC), E-38205 La Laguna, Tenerife, Spain}
\affiliation{Departamento de Astrof\'isica, Universidad de La Laguna (ULL), E-38206 La Laguna, Tenerife, Spain}

\author[0000-0002-3940-2360]{Don J. Radford} 
\affiliation{American Association of Variable Star Observers, 49 Bay State Road, Cambridge, MA 02138, USA}

\author{Howard M. Relles}
\affiliation{Center for Astrophysics \textbar \ Harvard \& Smithsonian, 60 Garden Street, Cambridge, MA 02138, USA}

\author{Gregor Srdoc}
\affil{Kotizarovci Observatory, Sarsoni 90, 51216 Viskovo, Croatia}

\author[0000-0003-2163-1437]{Chris Stockdale}
\affiliation{Hazelwood Observatory, Australia}

\author[0000-0001-5603-6895]{Thiam-Guan Tan}
\affiliation{Perth Exoplanet Survey Telescope, Perth, Western Australia}

\author[0000-0003-3092-4418]{Gavin Wang}
\affiliation{Tsinghua International School, Beijing 100084, China}


\begin{abstract}

The effect of stellar multiplicity on planetary architecture and orbital dynamics provides an important context for exoplanet demographics. We present a volume-limited catalog up to 300 pc of 66 stars hosting planets and planet candidates from \kepler, \textit{K2} and \TESS  with significant Hipparcos-Gaia proper motion anomalies, which indicates the presence of companions. We assess the reliability of each transiting planet candidate using ground-based follow-up observations, and find that the TESS Objects of Interest (TOIs) with significant proper anomalies show nearly four times more false positives due to Eclipsing Binaries compared to TOIs with marginal proper anomalies. In addition, we find tentative evidence that orbital periods of planets orbiting TOIs with significant proper anomalies are shorter than those orbiting TOIs without significant proper anomalies, consistent with the scenario that stellar companions can truncate planet-forming disks. Furthermore, TOIs with significant proper anomalies exhibit lower Gaia differential velocities in comparison to field stars with significant proper anomalies, suggesting that planets are more likely to form in binary systems with low-mass substellar companions or stellar companions at wider separation. Finally, we characterize the three-dimensional architecture of LTT 1445 ABC using radial velocities, absolute astrometry from Gaia and Hipparcos, and relative astrometry from imaging. Our analysis reveals that LTT 1445 is a nearly flat system, with a mutual inclination of $\sim2.88^{\circ}$ between the orbit of BC around A and that of C around B. The coplanarity may explain why multiple planets around LTT1445 A survive in the dynamically hostile environments of this system.

\end{abstract}


\section{Introduction} \label{sec:intro}

Radial velocity (RV, \citealt{Cumming2008, Fulton2021}) surveys and space-based transit searches such as \kepler \citep{Borucki2010, HowardMarcy2012} and \TESS \citep{Ricker2014} have revolutionized our understanding of exoplanet demographics. However, the process of confirming exoplanets is biased against stars in multiple systems, since close companions complicate the observations and analysis. Although one third of nearby solar-type stars have at least one companion \citep{Raghavan2010}, the effects of stellar multiplicity on planetary architecture and orbital dynamics are still poorly understood. In addition, unknown stellar companions can cause inaccuracy for estimating planet radius by diluting the measured transit depths \citep{Furlan2017, Teske2018,Sullivan2023}. Planet properties may also be inaccurate if the planet is actually orbiting the secondary star. Thus, identifying the stellar companions of transiting planets helps to obtain more accurate planet parameters and characterize the demographics of planets in binaries \citep{Fontanive2021, Cadman2022}. By analyzing large samples of planets in binaries and comparing them to single systems, we can gain insights into the factors that shape the the formation and evolution of exoplanets.

Close companions are expected to have a deleterious influence on planet formation, through disk truncation \citep{AandL1994,Jang-Condell2015} or dynamical stirring of planetesimals \citep{Quintana2007}. Recent ALMA observations show that disks in binaries have lower masses \citep{Akeson2019} and smaller radii \citep{Cox2017,Manara2019}, supporting the disk truncation scenario. \citet{Kraus2016} used high-resolution adaptive optics (AO) imaging of 382 Kepler Objects of Interest (KOIs) to show that planet occurrence rate in close binaries ($< 47 $ AU) is only 0.34 times that of single stars or wide binaries. \citet{Ziegler2020, Ziegler2021, Howell2021, Lester2021} performed similar searches for stellar companions to TESS Objects of Interest (TOIs) and also found a deficit of close binaries ($<100 $ AU). For giant planets discovered by RV observations, \citet{Hirsch2021} reported that the planet occurrence rate in binaries with a separation $< 100$ AU is significantly smaller than those in binaries with a separation $> 100$ AU or single stars. Additionally, \cite{Fontanive2021} presents a volume-limited sample of companions from tens of AU out to 20000 AU in the literature and Gaia DR2 to exoplanet host stars. They found giant planets with masses above 0.1 $M_{J}$
are more frequently seen than small sub-Jovian planets in binary systems, which is supported by the simulations from Cadman et al. 2022.

However, some planets survive in such dynamically challenging environments for reasons that are still unclear \citep[]{Hatzes2003,Correia2008,Kane2015,Dupuy2016}. Close binary companions could induce gravitational perturbations on their orbits, causing the migration or spin-orbit misalignment of planets. Furthermore, close companions may torque the protoplanetary disks where planets form, and therefore shape the architecture of the planet systems. Studying the architecture of planets in close binary systems will shed light on the planet formation and evolution in these systems. 


Transit surveys including \textit{TESS}, \textit{Kepler} and \textit{K2} offer an unbiased planet sample in terms of stellar multiplicity. The coarse spatial resolution of \textit{TESS} ($21^{\arcsec}/pixel$) and \textit{Kepler} ($4^{\arcsec}/pixel$) makes it essential to conduct ground-based follow-up observations to resolve close binary systems. Previous studies have used adaptive optics (AO) and speckle imaging to search for stellar companions to planet candidate hosts discovered by 
\kepler and \TESS missions \citep{Kraus2016, Ziegler2020, Ziegler2021, Howell2021, Lester2021}. \Gaia mission's Renormalised Unit Weight Error (RUWE) is also an indicator for companions, as RUWE values are sensitive to the deviation from the single-star astrometic model. However, RUWE values are most effective for detecting binaries with separations from $0^{\arcsec}.1$ to $0^{\arcsec}.6$ \citep{Lindegren2018, Ziegler2020}. The companions located outside of the range might be overlooked.

In this paper, we use \Gaia \citep{Gaia} and \HIP \citep{ESA} proper motion anomalies \citep{Kervella2019, Brandt2019a} to identify close companions hiding in the large pixels of \textit{TESS} or \textit{Kepler}. The method takes advantage of $\sim25$ years time baseline between the two missions, and is sensitive to companions with orbital periods from decades to centuries \citep{Kervella2019}. Furthermore, the combination of \Gaia and \HIP astrometry, radial velocities (RVs) and imaging astrometry makes it possible to determine the 3D orbits of the companions and obtain their dynamical masses \citep{Brandt2019a,Xuan2020}. In this paper, we characterized the 3D orbits of companions in the proof-of-concept system  LTT 1445 ABC with the method. It's important to note that our emphasis here is on the orbit parameters of stellar companions with orbital periods in years, rather than on transiting planets with much shorter orbital periods in the order of days.  Finally, By obtaining the inclinations of the companion orbits, we can constrain the mutual inclinations between the orbital plane of the companion and that of the transiting planet. This information can provide insight into the system's dynamical history.

\section{Hipparcos-Gaia proper motion anomalies}\label{sec:HGA}

The \textit{Gaia} spacecraft \citep{Gaia} measures the position and proper motion of nearly 1.7 billion stars since 2014. Its predecessor \textit{Hipparcos} \citep{Perryman1997} also provides precise astrometric measurements of nearby stars from 1989 to 1993. The measurements have a time baseline of nearly 25 years and can detect the effect of unresolved binaries since a companion would cause the primary to wobble around the barycenter on the sky plane \citep{Brandt2019a, Kervella2019}. Specifically, we use \HIP and \Gaia EDR3 proper motions and their uncertainties from the Hipparcos-Gaia Catalog of Accelerations (HGCA, \citealt{Brandt2021}). The catalog provides three proper motions for every star: (1) the \HIP proper motion $\mu_{\rm{H}}$ at an epoch near 1991.25; (2) the \Gaia EDR3 proper motion $\mu_{\rm{G}}$ at an epoch near 2016.01; (3) long-term proper motion $\mu_{\rm{HG}}$ given by position difference between \HIP and \Gaia  divided by the $25$ years baseline.

These proper motions are in units of \masyr. The proper motions are given in right ascension (RA, $\alpha$) and declination (DEC, $\delta$) direction. For simplicity, we use the total proper motion combined from the two directions and omit the subscript for RA and DEC in this paper. The long-term proper motion $\mu_{\rm{HG}}$ can be used to estimate the velocity of celestial linear motion across the sky plane over nearly 25 years. We subtracted the long-term proper motion $\mu_{\rm{HG}}$ from \HIP  and \Gaia proper motions as follows:

\begin{equation}
    \begin{aligned}
        \Delta \mu_{\rm{H}} & = \mu_{\rm{H}} - \mu_{\rm{HG}}\\
        \Delta \mu_{\rm{G}} &= \mu_{\rm{G}} - \mu_{\rm{HG}}
    \end{aligned}
\end{equation}
The two residuals represent the proper motion anomalies at the \HIP and \Gaia epochs, respectively \citep{Kervella2019}. Note that these anomalies are also known as astrometric accelerations \citep{Brandt2021}. We use the terminology of "proper motion anomalies" to prevent confusion with the general concept of "acceleration," as the residuals are measured in the unit of $\rm{mas\ year^{-1}}$.

As shown in Figure~\ref{fig:figure1_2}, a significant proper motion anomalies  reveals a deviation from the linear stellar motion, possibly caused by a gravitationally bound companion. We calculate the signal-to-noise ratio of proper motion anomalies  at \HIP and \Gaia epochs using the calibrated uncertainties of \HIP and \Gaia measurements from \cite{Brandt2021}: 
\begin{equation}\label{eqa:snr}
    \begin{aligned}
        \rm{SNR_{G}} =   \frac{\mu_{G}-\mu_{HG}}{\sqrt{\sigma [\mu_{G}]^{2}+\sigma [\mu_{HG}]^{2}}} \\
        \rm{SNR_{H}} =   \frac{\mu_{H}-\mu_{HG}}{\sqrt{\sigma [\mu_{H}]^{2}+\sigma [\mu_{HG}]^{2}}}
    \end{aligned}
\end{equation}
where $\sigma[ \mu]$ represent the uncertainties.

Next, we convert the proper motion anomalies  in the unit of $mas\ yr^{-1}$ into the differential velocities in the unit of $m\ s^{-1}$ as follows \citep{Kervella2019}:
\begin{figure}
    \centering
    \includegraphics[width=\linewidth]{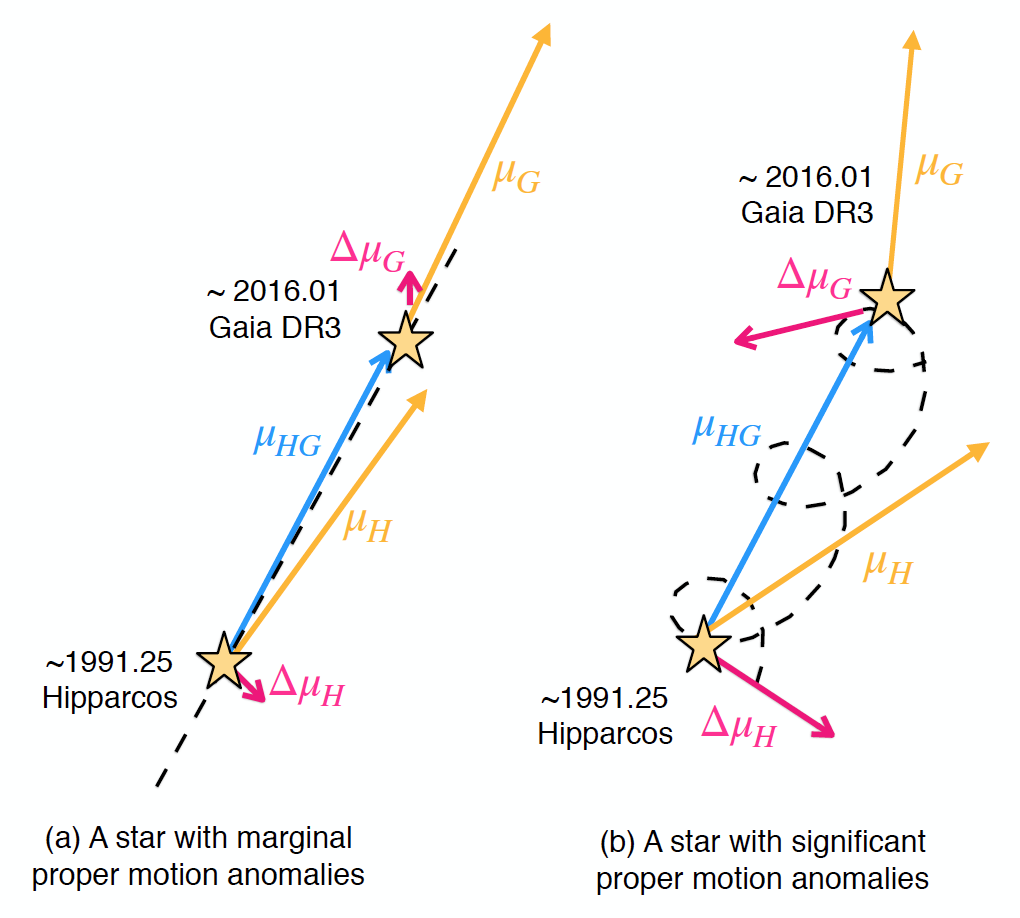}
    \caption{Principle of Hipparcos Gaia  proper motion anomalies. $\mu_{\rm{H}}$ and  $\mu_{\rm{G}}$ represent the proper motions of the same star measured by Hipparcos and Gaia with a time baseline of around 25 years. $\mu_{\rm{HG}}$  is the stellar long-term velocity across the sky plane. If we subtract  $\mu_{\rm{HG}}$ from  $\mu_{\rm{H}}$ and  $\mu_{\rm{G}}$, the residuals are the proper motion anomalies at Hipparcos and Gaia epochs, respectively.  (a)  A star with marginal proper motion anomalies: if  $\mu_{\rm{H}}$ and $\mu_{\rm{G}}$ are similar to the long-term velocity $\mu_{\rm{HG}}$, the star moves across the sky plane in a linear motion. (b) A star with significant proper motion anomalies: a significant residual indicates the star not only moves linearly but also orbits around the system barycenter due to the gravitational pull from a companion.   }
    \label{fig:figure1_2}
\end{figure}

\begin{equation}\label{eqn:3}
    \begin{aligned}
        \Delta v_{G}[m\ s^{-1}] = \frac{\Delta \mu_{G}[mas\ yr^{-1}]}{\varpi[mas]}\times 4740.47\\
        \Delta v_{H}[m\ s^{-1}] = \frac{\Delta \mu_{H}[mas\ yr^{-1}]}{\varpi[mas]}\times 4740.47
    \end{aligned}  
\end{equation}

where $\varpi$ is the parallax in units of $mas$. For a binary system, the differential velocity is approximately the projected tangential velocities of the primary's orbital motion on the sky plane \citep{Kervella2019}. Based on Kepler's law, the differential velocities are proportional to the companion masses ($m_{c}$) and inversely proportional to the square root of orbital distances ($r$): $\Delta v \propto \frac{m_{c}}{\sqrt{r}}$. Due to the observing window smearing effect (for details see \citealt{Kervella2019}), the proper motion anomalies  method is most sensitive to companions with orbital periods longer than observing windows of \HIP and \Gaia ($\delta_{H}=1227$ days, \citealt{Perryman1997}, $\delta_{\rm{G, DR3}}=1038$ days, \citealt{Gaia}). On the other hand, the efficiency of the proper motion anomalies  method drops for companions at orbital periods much longer than the  25-year baseline between the two missions. For instance, the efficiency is reduced to $\sim 30\%$ when the orbital period is ten times the time baseline ($\sim 250$ years) \citep{Kervella2019}. Therefore, the sweet spot of proper motion anomalies  is for companions at orbital periods from $\sim 3 $ years up to $\sim 250$ years, corresponding to a few AU to dozens of AU in terms of the semi-major axis. Multiple studies have found a deficiency of planets in close binaries with separation below 100 AU, supporting the theory of close companions disturbing and preventing planet formation \citep{Kraus2016, Ziegler2020, Hirsch2021,Fontanive2021,Cadman2022}. \HIP and \Gaia astrometry thus offers an efficient way to search for planets in binaries that have separation of $<100$ AU, with which we can study the effect of companions on planet formation and evolution.  


\section{Target Selection}\label{sec:TS}

\subsection{Methodology}
We constructed our target sample from host stars of transiting planet candidates, including TESS/Kepler Object of Interest (TOIs/KOIs) and K2 planet candidates. We used 4763 KOIs (2402 confirmed planets) from \kepler \citep{Batalha2013, Burke2014,Rowe2015, Mullally2015,Coughlin2016, Thompson2018}, 1547 K2 planet candiates (569 confirmed planets) from \textit{K2} mission \citep{Howell2014,Huber2016, Pope2016,Kostov2019,Zink2021}, and 6682 TOIs (360 confirmed planets) from \TESS \citep{Guerrero2021}. We downloaded the TOI/KOI/K2 lists from NASA Exoplanet Archive\footnote{https://exoplanetarchive.ipac.caltech.edu/}.  Next, we used the Hipparcos and Gaia proper motion anomalies  as an indicator to search for hidden companions to the planet or planet candidate hosts. Our procedure for selecting targets is as follows:
\begin{figure}
    \centering
    \includegraphics[width=\linewidth]{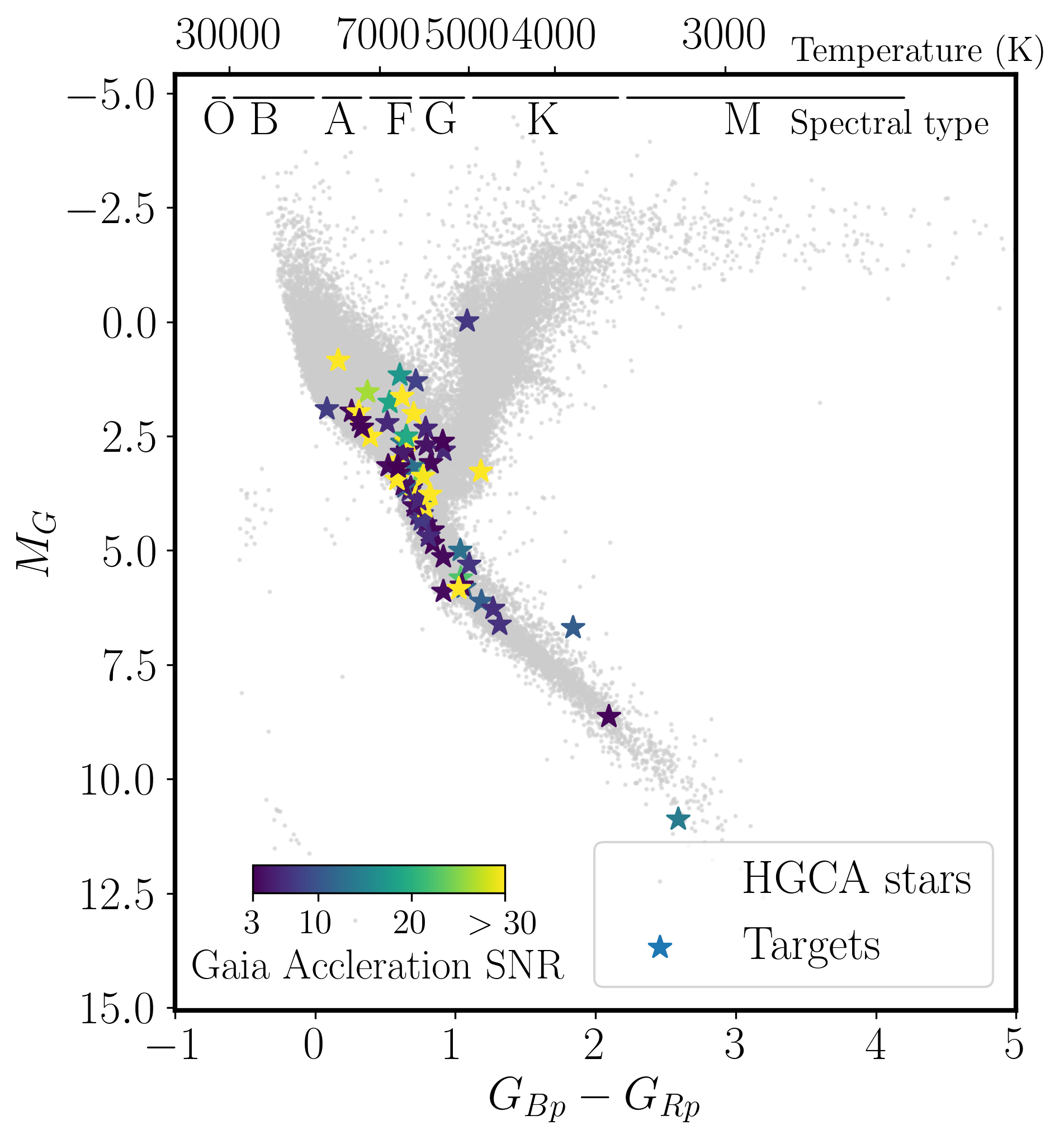}
    \caption{ Gaia color-absolute magnitude diagram for our targets (star signs) and stars in HGCA within 300 pc (grey dots). TOI/KOI/K2 planet candidate hosts with significant proper motion anomalies are color-coded by the signal-to-noise ratio of their proper motion anomalies at Gaia epoch ($\rm{SNR_{G}}$).}
    \label{fig:figure1}
\end{figure}
\begin{enumerate}

    \item We cross-matched KOI/K2/TOI lists with HGCA using their RA ($\Delta \alpha<10\arcsec$), DEC ($\Delta \delta<10\arcsec$) , and parallax ($\Delta \varpi / \varpi_{\rm{Gaia}} < 20\%$). 
    \item We calculated the distance of targets with \Gaia DR3 parallax and selected stars with distances smaller than 300 pc. 
    \item We calculated the signal-to-noise of \Gaia proper motion anomalies  of every star from the last step. We selected those showing \Gaia proper motion anomalies  with $\geq 3\sigma$ significance ($\rm{SNR_{G}} \geq 3$) into our target sample. Hereafter, we refer to our sample as HGCA-high-SNR stars or stars with significant proper motion anomalies.
    \item We also constructed a control sample of TOI stars with $\rm{SNR_{G}} <3$. Hereafter, we refer to the control sample as HGCA-low-SNR stars or stars with marginal proper motion anomalies. 
\end{enumerate}

There are a total of 66 systems (58 TOIs, 4 KOIs, and 4 K2 planet candidates) in our target sample with high-SNR proper motion anomalies  (see Table~\ref{tab:table1}).We also identified 254 TOIs with low-SNR proper motion anomalies  in the control sample and list them in Appendix~\ref{sec:toilow}. The HGCA provides a parameter called $\chi^{2}$, which represents the chi-squared value obtained from fitting a constant proper motion to the more precise pair of $\mu_{\rm{HG}}-\mu_{\rm{G}}$ and $\mu_{\rm{H}}-\mu_{\rm{HG}}$ proper motion measurements \citep{Brandt2021}. This parameter is also helpful in evaluating the significance of proper motion anomalies. For instance, a $\chi^{2}$ value of 11.8 corresponds to a $3 \sigma$ evidence for proper motion anomalies. We compare the output samples using criteria of  $\chi^{2}\geq 11.8$ and $\rm{SNR_{G}\geq 3}$. The two criteria are nearly equvalent, resulting in comparable samples.

\subsection{Target Sample}

 Figure~\ref{fig:figure1} presents a color-absolute magnitude diagram of 66 selected targets (58 TOIs, 4 KOIs and 4 K2 planet candidates), color-coded by the significance of Hipparcos-Gaia proper motion anomalies. We calculated the absolute magnitude using $M_{G}=m_{G}+5\log{d}+5$, where $m_{G}$ and $d$ are Gaia G band apparent magnitude and distance. Most of our targets are main sequence stars with $0.5<G_{Bp}-G_{Rp}<1.5$ and $0<M_{G}<7$. We highlight stars with multiple transiting planets in Table~\ref{tab:table1}, including TOI-402 \citep{GD2019,DX2019}, TOI-455 (LTT 1445 \citealt{winters2019,Winters2022}), TOI-144 ($\pi$ Men \citealt{JH2002,GD2018, HAP2022}), TOI-201 \citep{HM2021}, TOI-1339 (HD 191939, \citealt{BAM2020,Lubin2022,OM2023}), TOI-1730 \citep{SACR2022}, and KOI03158 (Kepler-444, \citealt{Ca2015}). 

Some targets in our sample have been investigated before. The triple system Kepler-444 consists of a primary star with five transiting Mars-sized and Mars-mass planets. \cite{Dupuy2016} and \cite{Mills2017} characterized the Kepler-444 BC companion pair as orbiting the primary A in a highly eccentric orbit ($e\sim0.864$, $a\sim5$ AU) using RVs and relative astrometry from imaging. Recently, \cite{Zhangzj2022} improved the constraints on the orbit of the Kepler-444 BC pair ($e\sim0.55$, $a\sim36$ AU, $i\sim85.4$ deg) using a longer time baseline of RVs and the proper motion anomalies  data from the \HIP and \Gaia missions. Both studies suggest that Kepler-444 BC may have truncated the protoplanetary disk of the primary, resulting in the small sizes of the system's five planets. \cite{zhou2022} characterizes the 3D orbit of an M-dwarf companion to TOI-4399 (HIP 94235), which hosts a mini-Neptune with an orbital period of 7.1 days. Their results show that the companion has a semi-major axis of $\sim60$ AU and an inclination of $\sim 67.8^{\circ}$, indicating a modest misalignment between the companions and the transiting planet. Furthermore, previous studies have identified that the Hipparcos and Gaia proper motion anomalies  of TOI-144 ($\pi$ Men,  \citealt{Xuan2020,DR2020,Damasso2020}), TOI-1144 (HAPT-11, \citealt{Xuan2020}) and TOI-1339 (HD 191939,  \citealt{Lubin2022}) are from giant planets at a few AU. By combining the proper motion anomalies  and RVs, these studies obtained a constraint on the semi-major axis, orbital inclination and mass of the giant planets. 

 Among the sample of 66 targets with high proper motion anomalies, 33 systems have confirmed companions from previous surveys using AO/speckle imaging, mostly at separations from $0.1^{\arcsec}$ to $2^{\arcsec}$. We list the companion separations in Table~\ref{tab:table1} from published papers \citep{WDS2001,Kraus2016, Ziegler2020,Ziegler2021,winters2019, Howell2021, Lester2021} and TESS Follow-up Observing Program (TFOP). Additional AO imaging using Keck/NIRC2 and Subaru/SCExAO will be presented in a follow-up paper.

\begin{center}
\begin{longtable*}{lccccccccc}
\caption{TOI/KOI/K2 with significant proper motion anomalies}\label{tab:table1}\\
\hline
Name & HIP & $P_{pl}$ & Exoplanet Archive & This work&$\Delta v_{G}$ &  $\rm{SNR_{G}}$ & RUWE & distance$^{\rm{d}}$ & comp. sep.$^{\rm{e}}$\\
 &  Number &  days&  Disposition &  Disposition & $m\ s^{-1}$ & &  & pc & arcsec  \\
\hline
\hline
\endfirsthead
\hline
Name & HIP & $P_{pl}$& Exoplanet Archive &This work &$\Delta v_{Gaia}$ &  $\rm{SNR_{G}}$ & RUWE & distance& comp. sep.$^{\rm{d}}$ \\
 &  Number &  days&  Disposition & Disposition & $m\ s^{-1}$ & &   & pc & arcsec \\
\hline
\hline
\endhead
\hline
\endfoot
TOI 1684 & 20334 & 1.16 & PC& PC & 4176.05 & 226.52 & 1.58 & 87.11 & — \\
TOI 510 & 33681 & 1.35 & APC& PC &2700.32 & 116.11 & 2.43 & 92.84 & $5.5^{5}$ \\
TOI 394 & 15053 & 0.39 & APC& EB &6995.08 & 94.97 & 3.18 & 141.7  & $3.22^{1}$\\
TOI 6260 & 45961 & 2.39 & PC & PC & 6370.56 & 93.02 & 4.5 & 115.75 & — \\
TOI 271 & 21952 & 2.48 & APC& PC &8067.66 & 79.79 & 8.47 & 99.93  & $0.146^{2}$\\
TOI 1124 & 98516 & 3.52 & APC& EB &6904.05 & 53.18 & 11.77 & 80.17 & — \\
TOI 896 & 28122 & 3.53 & FA& FA & 3091.52 & 45.59 & 1.4 & 155.53 & $0.062^{2}$\\
TOI 5811 & 102295 & 6.25 & PC& PC & 2749.69 & 34.99 & 1.84 & 167.79 & —\\    
TOI 1418 & 83168 & 0.68 & FA& SV & 812.81 & 30.36 & 0.92 & 159.34  & — \\
TOI 953 & 21000 & 2.97 & FP& EB & 6963.46 & 30.07 & 0.93 & 202.99 & $4.42^{1}$\\
TOI 4314 & 22084 & 73.58 & FA& SV & 1370.99 & 29.95 & 2.23 & 153.05  & — \\
TOI 680 & 58234 & 0.43 & APC& PC & 1590.25 & 29.68 & 3.71 & 160.05 & $0.78^{1}$ \\
TOI 2118 & 79105 & 2.34 & FP& NEB & 2493.51 & 27.9 & 4.37 & 188.03 & — \\
TOI 6246 & 117249 & 6.78 & PC & PC &832.73 & 26.3 & 1.07  & 188.45 & - \\
TOI 1204 & 55069 & 1.38 & PC&  PC & 418.69 & 23.51 & 0.98 & 106.45 & $0.375^{2}$ \\
$\rm{TOI\ 402^{a}}$ & 11433 & 4.76 &  CP& CP & 187.62 & 22.46 & 0.89 & 44.86 & $1.44^{1}$ \\
$\rm{TOI\ 402^{a}}$ & 11433 & 17.18 &  CP& CP & 187.62 & 22.46 & 0.89 & 44.86 & $1.44^{1}$ \\
TOI 1946 & 70833 & 10.85 & FP& NEB & 3949.15 & 18.81 & 5.88 & 250.89 & — \\
TOI 930 & 16881 & 4.9 & PC& PC & 1673.63 & 17.24 & 1.43 & 294.7 & $0.69^{1}$ \\
$\rm{TOI\ 455^{a}}$ (LTT 1445)  & 14101 & 3.12 & CP& CP & 354.41 & 14.1 & 1.07 & 6.86 & A-BC $7.02^{4}$ \\
$\rm{TOI\ 455^{a}}$ (LTT 1445)  & 14101 & 5.36 & CP& CP & 354.41 & 14.1 & 1.07 & 6.86 & A-BC $7.02^{4}$ \\
TOI 4568 & 77921 & 14.01 & PC& PC& 309.25 & 13.4 & 6.3 & 48.29 & — \\
TOI 1665 & 27844 & 1.76 & FP& NEB & 126.38 & 13.17 & 0.94 & 66.44  & $1.9^{5}$ \\
TOI 1837 & 67650 & 5.82 & APC& EB & 1145.74 & 12.85 & 7.37 & 152.67  & $0.15^{3}$ \\
$\rm{TOI\ 2666^{b}}$ & 45621 & -- & APC& EB & 88.15 & 12.83 & 1.46 & 32.28 & — \\
TOI 179 & 13754 & 4.14 & PC&  PC& 72.24 & 12.06 & 0.99 & 38.63 & — \\
TOI 1099 & 108162 & 6.44 & PC& PC & 51.94 & 11.16 & 1.16 & 23.64 & $7.64^{1}$ \\
TOI 5521 & 57386 & 18.51 & PC&  PC & 420.92 & 10.78 & 1.68 & 136.95 & $0.6^{5}$ \\
TOI 1719 & 44289 & 2.75 & PC&  PC & 1864.24 & 8.31 & 10.38 & 236.54 & $0.12^{2}$ \\
TOI 1151 (KELT-20) & 96618 & 3.47 & CP& CP & 177.61 & 7.76 & 0.95 & 136.98 &  — \\
$\rm{TOI\ 144^{a}}$ ($\pi$ Men) & 26394 & 6.27 & CP& CP & 54.31 & 7.66 & 0.81 & 18.29 & $\rm{GP}^{6}$  \\
$\rm{TOI\ 144^{a}}$ ($\pi$ Men) & 26394 & 124.46 & CP& CP & 54.31 & 7.66 & 0.81 & 18.29 & $\rm{GP}^{6}$ \\
TOI 4399 & 94235 & 7.71 & CP& CP & 66.96 & 7.4 & 1.07 & 58.55 &$0.60^{8}$ \\
TOI 4175 & 64573 & 2.16 & PC& PC  & 131.36 & 7.18 & 1.34 & 46.32 & $1.67^{1}$ \\
TOI 2299 & 93711 & 165.02 & PC& PC & 74.33 & 6.87 & 1.06 & 34.27  &  — \\
TOI 1144 (HAPT-11) & 97657 & 4.89 & KP& CP& 42.03 & 6.52 & 0.95 & 37.84 & $\rm{GP}^{6}$ \\
$\rm{TOI\ 201^{a}}$ & 27515 & 5.85& PC& PC & 133.84 & 6.28 & 1.1 & 112.18 &  — \\
$\rm{TOI\ 201^{a}}$ & 27515 & 52.98 & CP& CP & 133.84 & 6.28 & 1.1 & 112.18 &  — \\
TOI 635 & 47990 & 0.49 & FA& PC & 85.89 & 6.22 & 1.03 & 59.04 & $1.77^{1}$ \\
TOI 1831 & 65205 & 0.56 & PC& PC & 179.31 & 6.08 & 0.96 & 126.45 & $0.65^{3}$ \\
TOI 1131 & 81087 & 0.59 & PC& PC &1835.93 & 6.06 & 12.51 & 247.64 & $0.11^{2}$ \\
TOI 4603 & 26250 & 7.25 & CP& CP & 307.19 & 5.92 & 1.0 & 224.15 &  — \\
TOI 909 & 82032 & 3.9 & FP& NEB & 42.47 & 5.81 & 1.04 & 51.65 & $1.35^{1}$ \\
TOI 575 & 41849 & 19.37 &  FP& EB & 161.45 & 4.96 & 1.02 & 178.49 & $0.63^{1}$ \\
TOI 141 & 111553 & 1.01 &  CP& CP & 37.88 & 4.92 & 1.09 & 47.77 &  0.44/$1.20 ^{1}$ \\
TOI 128 & 24718 & 4.94 & PC& PC & 63.49 & 4.8 & 0.94 & 68.23 & $2.22 ^{1}$\\
TOI 2017 & 74685 & 5.5 & FP& EB & 278.86 & 4.37 & 7.12 & 98.57 & —  \\
TOI 6265 & 49508 & 5.79 & PC & PC & 88.11 & 4.15 & 1.09 & 71.29 & -\\
TOI 5140 & 42403 & 15.61 & PC& PC & 24.67 & 3.85 & 0.98 & 35.24 & — \\
TOI 5962 & 112486 & 1.92 & PC & PC & 147.97 & 3.80 & 0.97 & 127.27 & — \\
TOI 230 & 114003 & 13.34 & FA& PC & 106.85 & 3.78 & 1.01 & 141.62 & —\\
TOI 5383 & 69275 & 2.8 & PC& PC & 120.17 & 3.73 & 2.53 & 102.09 & — \\
$\rm{TOI\ 1339^{a}}$ & 99175 & 8.88 & CP& CP &32.34 & 3.73 & 1.05 & 53.49 & $\rm{GP}^{6}$  \\
 $\rm{TOI\ 1339^{a}}$ & 99175 & 28.58 & CP& CP &32.34 & 3.73 & 1.05 & 53.49 & $\rm{GP}^{6}$  \\
  $\rm{TOI\ 1339^{a}}$ & 99175 & 38.35 & CP& CP &32.34 & 3.73 & 1.05 & 53.49& $\rm{GP}^{6}$  \\
   $\rm{TOI\ 1339^{a}}$ & 99175 &101.12 &CP& CP &32.34 & 3.73 & 1.05 & 53.49 & $\rm{GP}^{6}$ \\
    $\rm{TOI\ 1339^{a}}$ & 99175 & 284 & CP& CP &32.34 & 3.73 & 1.05 & 53.49 & $\rm{GP}^{6}$ \\
TOI 621 & 44094 & 3.11 & APC& PC & 97.46 & 3.69 & 0.84 & 182.31 & $2.06 ^{1}$ \\
TOI 5079 & 24007 & 1.49 & FP& PC & 99.1 & 3.47 & 0.98 & 76.93 & $2.7 ^{5}$\\
TOI 1271 & 66192 & 6.13 & KP& CP & 53.15 & 3.47 & 0.96 & 92.04 & — \\
$\rm{TOI\ 222^{b}}$ & 118045 & -- & FP& EB & 66.43 & 3.46 & 1.78 & 85.42 & — \\
TOI 522 & 40694 & 0.4 & PC& PC  & 110.24 & 3.36 & 0.99 & 111.73 & $0.75 ^{1}$ \\
$\rm{TOI\ 1730^{a}}$ & 34730 & 2.16 &  PC& CP& 52.09 & 3.32 & 1.14 & 35.65   & — \\
$\rm{TOI\ 1730^{a}}$ & 34730 & 6.23&  PC& CP& 52.09 & 3.32 & 1.14 & 35.65  & — \\
$\rm{TOI\ 1730^{a}}$ & 34730 & 12.56 &  PC& CP& 52.09 & 3.32 & 1.14 & 35.65  & —  \\
TOI 1799 & 54491 & 7.09 & PC& PC & 40.65 & 3.31 & 1.03 & 62.23  & — \\
TOI 369 & 115594 & 5.46 & PC& PC & 276.13 & 3.21 & 1.36 & 227.75  & $0.69 ^{1}$ \\
TOI 906 & 78301 & 1.66 & PC& PC & 104.0 & 3.04 & 0.92 & 135.41 &  $1.35 ^{1}$ \\
$\rm{K03158^{a}}$ (Kepler444) & 94931 & 3.60 & CP & CP & 256.64 & 45.44 &  0.99 & 36.55 & $1.84^{7}$\\
$\rm{K03158^{a}}$ (Kepler444) & 94931 & 4.55 & CP & CP & 256.64 & 45.44 &  0.99 & 36.55 & $1.84^{7}$\\
$\rm{K03158^{a}}$ (Kepler444) & 94931 & 6.18 & CP & CP & 256.64 & 45.44 &  0.99 & 36.55 & $1.84^{7}$\\
$\rm{K03158^{a}}$ (Kepler444) & 94931 & 7.74 & CP & CP & 256.64 & 45.44 &  0.99 & 36.55 & $1.84^{7}$\\
$\rm{K03158^{a}}$ (Kepler444) & 94931 & 9.94 & CP & CP & 256.64 & 45.44 &  0.99 & 36.55 & $1.84^{7}$\\
K06139 & 94780 & 0.91 & FP &EB & 424.76 & 7.3 & 1.9  & 237.73 & — \\
K01924 & 96501 & 2.12 & FP& EB& 1003.70 & 4.32 & 12  & 106.51 & —\\
K06364 & 93954 & 1.54 & FP & EB &117.26 & 3.29 &  1.58 & 139.6 & — \\
EPIC204165788 & 80474 & 0.75 &  FP & EB  &2264.77 & 47.42 &  2.7 & 123.2 & —\\
EPIC201488365 & 54766 & 3.36 & PC& EB& 592.84 & 19.27 & 0.96  & 146.39 & —\\
EPIC212096658 & 41431 & 1.47 & PC& EB & 550.58 & 10.96 &  10 & 49.43 & —\\
EPIC204506777 & 78977 & 1.63 & FP&EB & 204.61 & 3.25 &  3.29 & 150.92 & — \\
\hline
\end{longtable*}
\tablenotetext{a}{Multiple planet systems.}
\tablenotetext{b}{Single transit planet candidate, no orbital period.}
\tablenotetext{c}{The flags come from NASA Exoplanet Archive. CP: Confirmed Planet, PC: Planet Candidate, APC: Ambiguous Planet Candidate, FP: False Positive, FA: False Alarm.  }
\tablenotetext{d}{Distances are from Gaia DR3 release \citep{Gaia}}
\tablenotetext{e}{Companion separation from published papers or TFOP. The references are: 1. \cite{Ziegler2020,Ziegler2021}; 2. \cite{Lester2021}; 3. TFOP, observation using Palomar/PHARO (PI: D. Ciardi); 4. \cite{Rodriguez2015, winters2019}; 5. WDS catalog \citep{WDS2001}; 6. GP: the proper motion anomalies of TOI144, TOI1144 and TOI1339 are from giant planets \citep{Xuan2020, DR2020, Damasso2020, Lubin2022}; 7. \cite{Dupuy2016, Zhangzj2022}. 8. \cite{zhou2022} }
\end{center}

\section{Statistical analysis of TOIs }

\subsection{Planet candidates vs. False positives} \label{subsec:FP}

\begin{figure*}
    \centering
    \includegraphics[width=\linewidth]{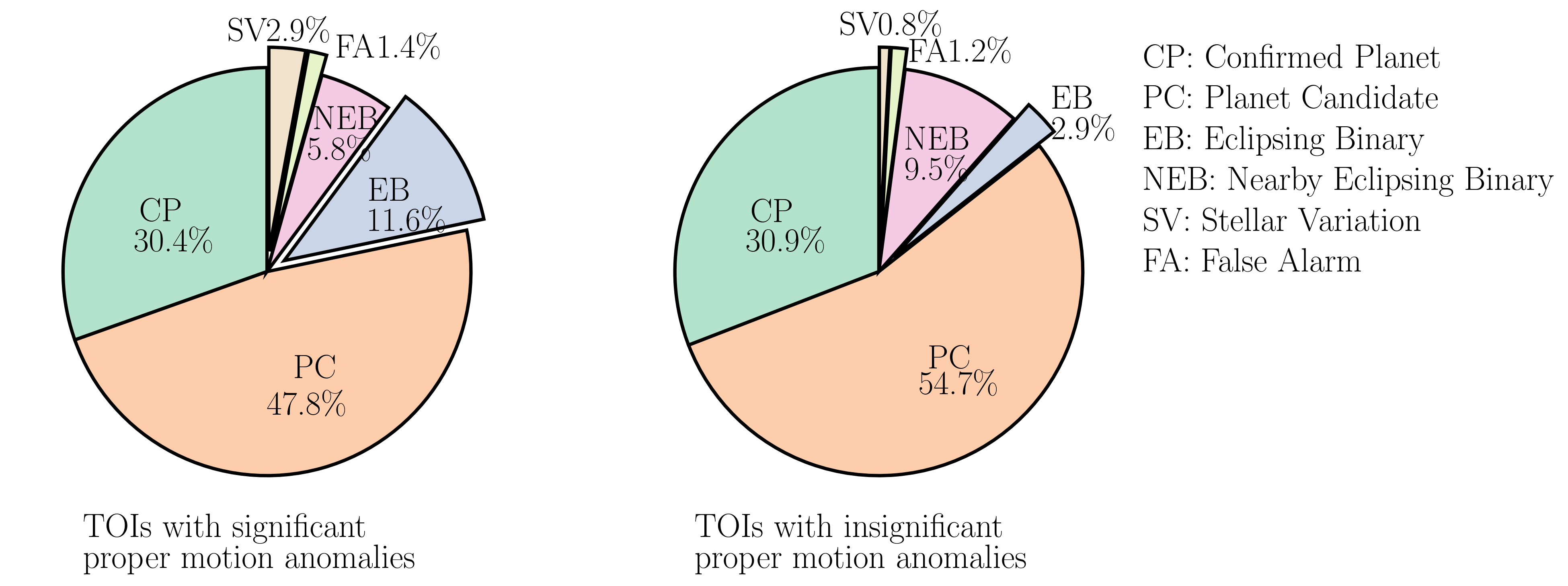}
    \caption{Right panel: the fraction of confirmed planets, planet candidates, and several false positives of TOIs in our high SNR sample. Left panel: the fraction of the same categories but for low SNR TOIs. The two samples are both limited to 300 pc. We only present TOIs in this figure because the majority of KOI/K2 targets in our sample are EBs.  }
    \label{fig:figure3}
\end{figure*}
We first analyzed the fraction of false positives of the transiting detections in the sample.
Table~\ref{tab:table1} lists the NASA Exoplanet Archive dispositions that classify TOIs/KOIs/K2 candidates, including Confirmed/Known Planet (CP/KP), Planet Candidate (PC), Ambiguous Planetary Candidate (APC), False Positive (FP), and False Alarm (FA).  We also present different reasons for the false positive/false alarm dispositions in Table~\ref{tab:table1}. A false positive or false alarm flag is assigned in several situations. The first scenario is \textit{Eclipsing Binaries} (EBs), in which the secondary stars graze the edge of primaries, and the reduction in brightness is indistinguishable from transits of smaller planets. The second scenario is the contamination by a \textit{Nearby Eclipsing Binary} (NEB) as multiple stars are unresolved due to the large pixel scale of \kepler and \TESS ($4^{\arcsec}$ for \kepler and $21^{\arcsec}$ for \TESS). In this case, the bright primary star dilutes the light of a nearby, dimmer, eclipsing binary pair to the point at which the eclipses seem as shallow as a planetary transit. In addition, stellar variation (SV) and spacecraft systematics errors (SSE) can also mimic the dips in light curves similar to those from transiting planets.

To break down the FP into the EB/NEB/SV flag, we refer to the TESS Follow-Up Observing Program Sub-Group 1$\&$2 (TFOP SG1 $\&$ SG2) disposition and notes as a guide for TOIs. TFOP SG1 performs seeing-limited imaging of the TOIs using ground-based telescopes with higher spatial resolution to check whether the transits occur on target. They detected four TOIs (TOI-2118, TOI-1665, TOI-909, TOI-1946) in our sample as NEB. In addition, TFOP SG2 identified 8 EBs based on odd-even transit difference (TOI-394, TOI-1124,TOI-575, ) or RVs from TRES+FIES (TOI-953, TOI-1837, TOI-2017, TOI-2666,TOI-222). In our sample, the majority of KOIs and K2 transiting signal are from EBs based on the results from Kepler/K2 EB catalogs \citep{Slawson2011, Armstrong2015, Rizzuto2017, Kruse2019}. We present the details of each FP in our sample in Appendix~\ref{sec:ap1}. 

The left panel in Figure~\ref{fig:figure3} presents the fraction of confirmed planets, planet candidates, eclipsing binaries, nearby eclipsing binaries, and other false positives in TOIs with significant proper motion anomalies (58 TOIs). We have chosen to only present the results of TOIs for a homogeneous comparison, because follow-up observations for candidates of TOI, KOI, and K2 are conducted through various projects, and the majority of KOIs and K2 targets in our sample are eclipsing binaries (EBs). For comparison, we show a control sample from TOIs with $\rm{SNR_{G}}<3$ and distances smaller than 300 pc (254 TOIs). We also break down the false positives in the HGCA-low-SNR sample into the same categories as our targets based on the TFOP SG1 disposition. The HGCA-high-SNR TOIs contain a higher fraction of false positives (up to $21.8\%$ compared to the $14.4\%$ of HGCA-low-SNRs TOIs). The difference is mainly from the excess of EBs among the HGCA-high-SNR TOIs, taking up $\sim11.6\%$ of the sample. The EB false positives result from contamination of triple systems with close-in eclipsing binaries. Due to the dilution of light curves by multiple sources in the same pixel, the transit depth appears comparable to planetary transits around single stars. Our finding agrees with previous studies. \cite{Ziegler2020} presents that hot Jupiters are more common in binaries with wide companions compared to field stars. But \cite{Ziegler2021} argues that these findings can be attributed to false positive contamination arising from tertiary companions to closely orbiting eclipsing binaries. In contrast, other false positives, including NEB, stellar variability, and spacecraft false alarms, account for a similar share in the two samples. 

\begin{figure*}
    \centering
    \includegraphics[width=\linewidth]{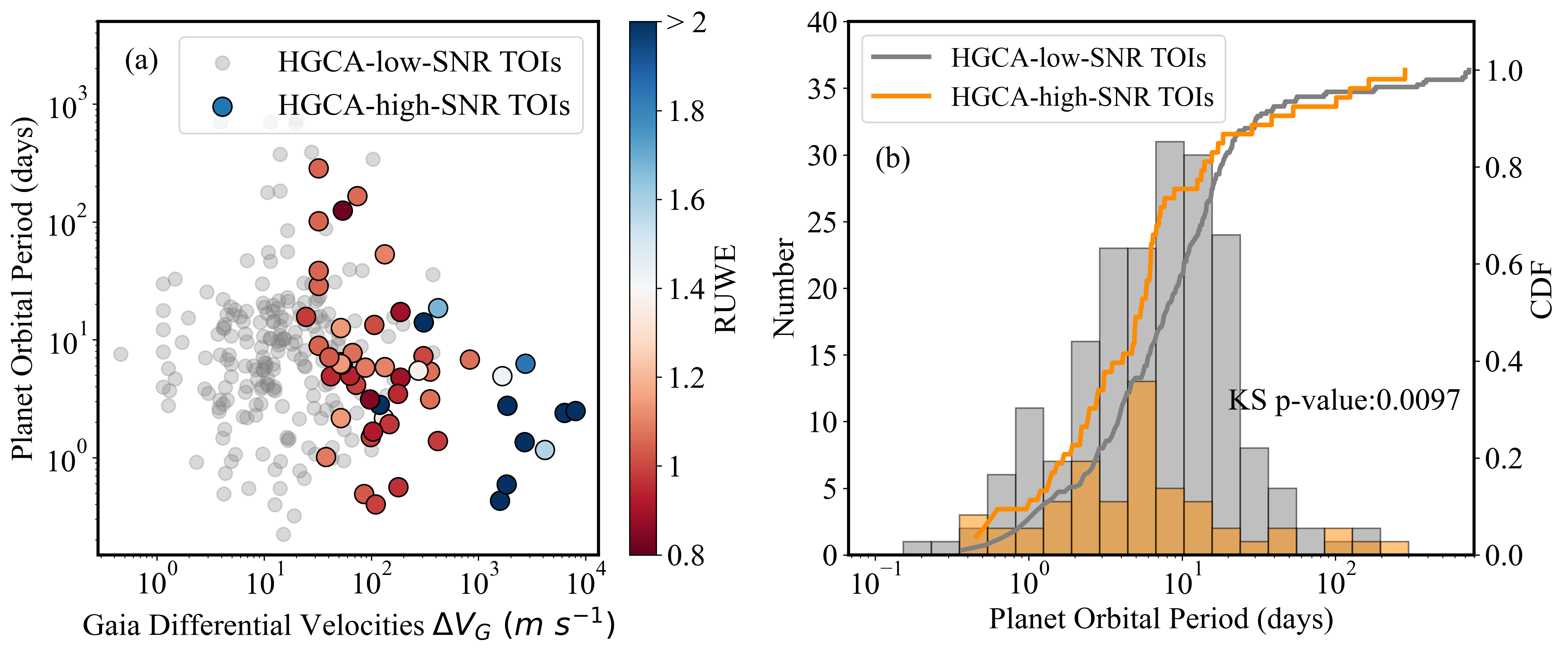}
    \caption{Panel a: planet orbital period vs. differential velocities at Gaia epoch of TOIs with significant proper motion anomalies (color-coded by their Gaia DR3 RUWE) and TOIs with marginal proper motion anomalies (grey). We exclude TOIs with FP/FA dispositions. High-SNR TOIs with RUWE $\ge 1.4$ are in red, whereas those with RUWE $<1.4$ are in blue. Panel b: marginalized distribution of planet orbital periods of high (orange) and low (grey) SNR TOIs overlapped with the cumulative distribution function. We exclude TOIs with FP/FA dispositions. }
    \label{fig:figure5}
\end{figure*}
\subsection{Orbital period of TOIs with significant proper motion anomalies}\label{sec:OPHA}

 In this section, we compare the orbital periods of planets around TOIs with significant Hipparcos-Gaia astrometric acceleration and those with marginal astrometric acceleration. Figure~\ref{fig:figure5}a presents the planet orbital periods and the Gaia differential velocities of TOIs with high-SNR and low-SNR proper motion anomalies, respectively. The Gaia differential velocities are defined in Section~\ref{sec:HGA} (see Eqn.~\ref{eqn:3}) We exclude systems with false positive and false alarm dispositions. We do not include the eight KOIs and K2 targets because seven of them are false positives. We colored HGCA-high-SNR TOIs by their Gaia DR3 Renormalised Unit Weight Error (RUWE). A RUWE value greater than 1.4 usually indicates that the source is non-single and the two components are too close to be fully resolved by Gaia \citep{Lindegren2018}. 

\begin{figure*}
    \centering
    \includegraphics[width=0.6\linewidth]{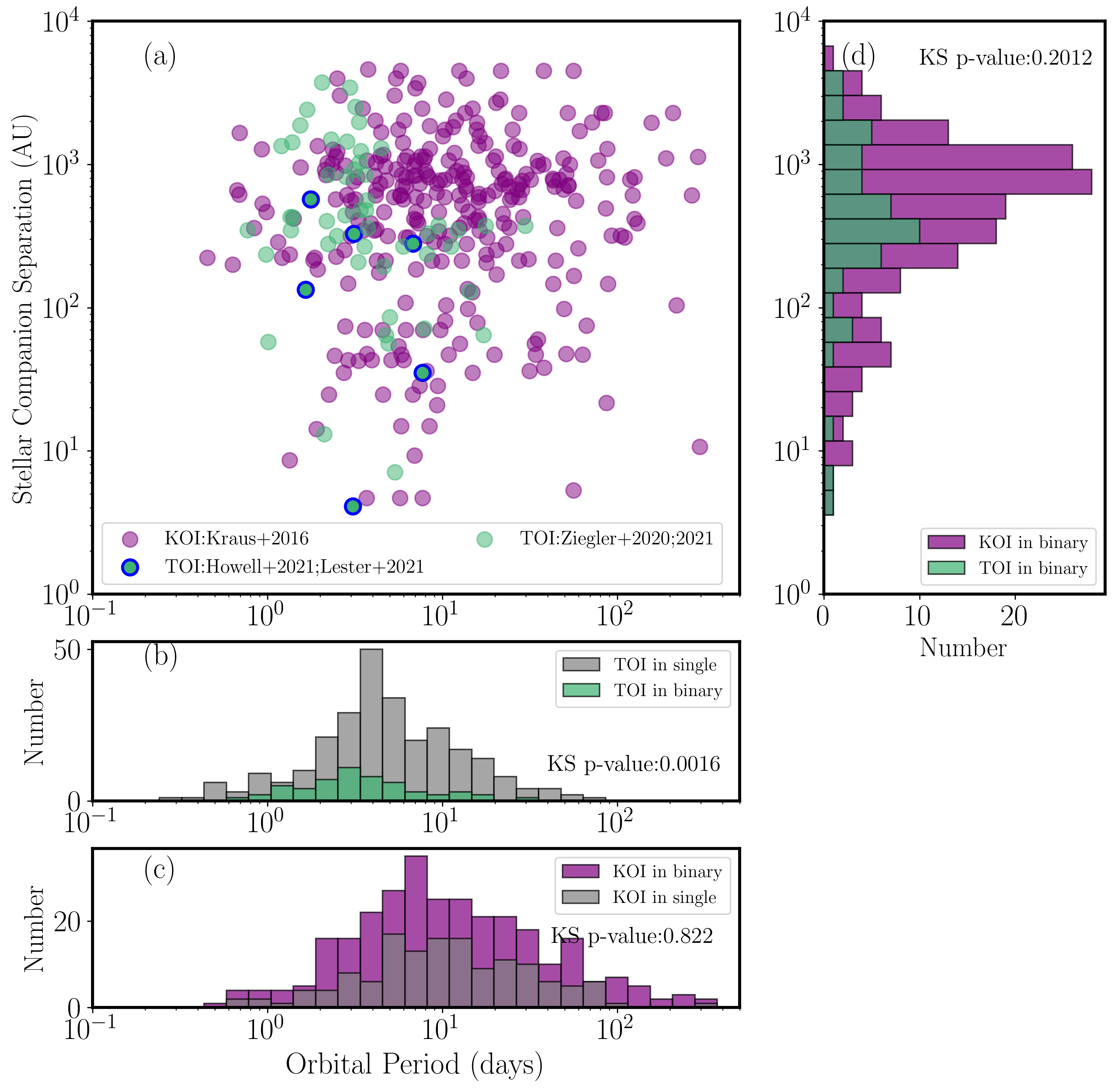}
    \caption{Panel a: planet orbital period vs. stellar companion separations of TOIs (green, \citealt{Ziegler2020,Ziegler2021,Lester2021,Howell2021}) and KOIs (purple,\citealt{Kraus2016}) in binary systems. We only include  confirmed planets (CP) or known planets (KP). Panel b: marginalized distribution of planet orbital periods of TOIs in binaries (green) and singles (grey). Panel c: same as panel b but for KOIs in binary (purple)  and single (grey) systems. Panel d: marginalized distribution of stellar companion separations of TOIs (green) and KOIs (purple) in binaries.}
    \label{fig:figure5n2}
\end{figure*}

The HGCA-high-SNR TOIs with high RUWE mostly have differential velocities beyond one thousand $m\ s^{-1}$, including TOI-271, TOI-680, TOI-930, TOI-1719, and TOI-1131. The high differential velocities are consistent with these TOIs having close stellar companions with separations below $1^{\arcsec}$ (see Table~\ref{tab:table1})\footnote{Except for TOI-510 which has a companion at $5.5^{\arcsec}$ reported from WDS catalog. The Hipparcos Gaia acceleration indicates there might be another unresolved companion at a closer separation.}. In comparison, HGCA-high-SNR TOIs with low RUWE have differential velocities from dozens to a few hundred $m\ s^{-1}$. In some systems, the velocities in the middle range are from stellar companions at relatively wider separation. For instance, TOI-402, TOI-4175, TOI-635, TOI-128 are all in this regime and have companions with separations $>1^{\arcsec}$. In other cases, low-mass companions, such as brown dwarf companions and giant planets, cause the primaries to orbit around the barycenter at velocities from dozens to a few hundred $m\ s^{-1}$. Their low RUWE indicates that the single-star model is still a good fit because the substellar companions are much fainter than primaries. In our sample, the Hipparcos-Gaia proper motion anomalies  reveal the existence of giant planets at a few au in TOI-144($\pi$ Men), TOI-1144(HAPT-11), and TOI-1339 systems.

Figure~\ref{fig:figure5}a shows a tentative inverse correlation between the orbital period of transiting planets and the Gaia differential velocities of their host stars. The planet periods of TOIs with differential velocities $>1000\ m\ s^{-1}$ are all shorter than ten days. Figure~\ref{fig:figure5}b displays the distributions of planet orbital periods of HGCA-high-SNR and HGCA-low-SNR TOIs. We include 53 HGCA-high-SNR TOIs and 200 HGCA-low-SNR TOIs. We can see that orbital periods of transiting planets in HGCA-high-SNR TOIs are generally shorter than in HGCA-low-SNR TOIs. A Kolmogorov-Smirnov test shows that the difference between two samples' planet orbital period distributions is statistically significant (p-value $= 9.7 \times 10^{-3}$).

To explore whether the trend found in Figure~\ref{fig:figure5} hold true for a larger sample, we compare the orbital periods of planets in binary and single TOIs/KOIs from \cite{Kraus2016, Ziegler2020, Ziegler2021,Howell2021,Lester2021}. Figure~\ref{fig:figure5n2}a presents the stellar companion separations and planet orbital periods of these TOIs and KOIs. In Figure~\ref{fig:figure5n2}b, We include 265 and 56 confirmed planets in binary and single TOIs from \cite{Ziegler2020, Ziegler2021, Howell2021,Lester2021}. In Figure~\ref{fig:figure5n2}c, we include 138 and 296 confirmed planets in binary and single KOIs from \cite{Kraus2016}. To rule out the influence of EBs that usually have short periods, we only select confirmed planets (CP) or known planets (KP) from their sample. We also removed duplicate TOIs resulting from the overlap between different surveys.  We did not apply a distance cutoff to their sample, as such an approach would have resulted in a further decrease in the size of the sample. But most of their targets are within 400 pc, with a small fraction extending beyond 800 pc (See  Figure 1 in  \citealt{Kraus2016} and \citealt{Ziegler2021}).We find that TOIs in binaries have orbital periods statistically shorter than those in single systems (p-value $=0.0016$). However, we do not see a significant difference between confirmed Kepler planets in binaries and singles (p-value=0.82).

A few possibilities can explain the different  results in the TOI and KOI samples. First, the difference in sensitivity between the two missions may explain the observed disparity, as \TESS detected more short-period planets within ten days but fewer at longer periods than \textit{Kepler}. However, this explanation alone cannot account for the shorter orbital periods of planets in binary TOIs, since \TESS searches for planets without considering the binarity of the targets. The second possibility is that stellar companions in KOI sample have relatively larger separations as they are generally more distant and fainter than TOIs. Therefore, companions in KOI sample are less influential in shaping the planets' orbital periods. \cite{Ziegler2020, Ziegler2021} and \cite{Kraus2016} both present stellar companions to KOIs/TOIs at separation from a few AU to a few thousand AU. However, Figure~\ref{fig:figure5n2}d illustrates that the distributions of stellar companion separations in the TOI and KOI samples do not exhibit a significant difference (p-value=0.20). Thirdly, the difference between binary and single TOIs might potentially be attributed to the relatively small size of TOIs in binary systems. Therefore, a larger sample of TOIs in binaries is required to reach a more decisive conclusion. Finally, the disparity in planet periods observed between the TOI and KOI samples may be due to the fact that \textit{Kepler} has higher precision and is thus more sensitive to smaller planets than \TESS. Consequently, the two missions may be observing different populations of planets. To test the hypothesis, we need to revise the radii of planets orbiting TOIs/KOIs in binary systems by accounting for the flux dilution.

In short, the reason why the \TESS bias affects binary and single systems differently is not yet understood, and a larger sample of TOIs is needed to draw a more definitive conclusion. If the TOIs in binaries do have shorter orbital periods (<10 days), they might form in truncated disks by the companions. Besides, planets' survival probability is likely higher at close-in orbits because the host stars provide more shield to resist the gravitational disturbance from the companions.

\subsection{Differential velocity distribution of TOIs vs. field stars}\label{sec:DVFS}

\begin{figure}
    \centering
    \includegraphics[width=\linewidth]{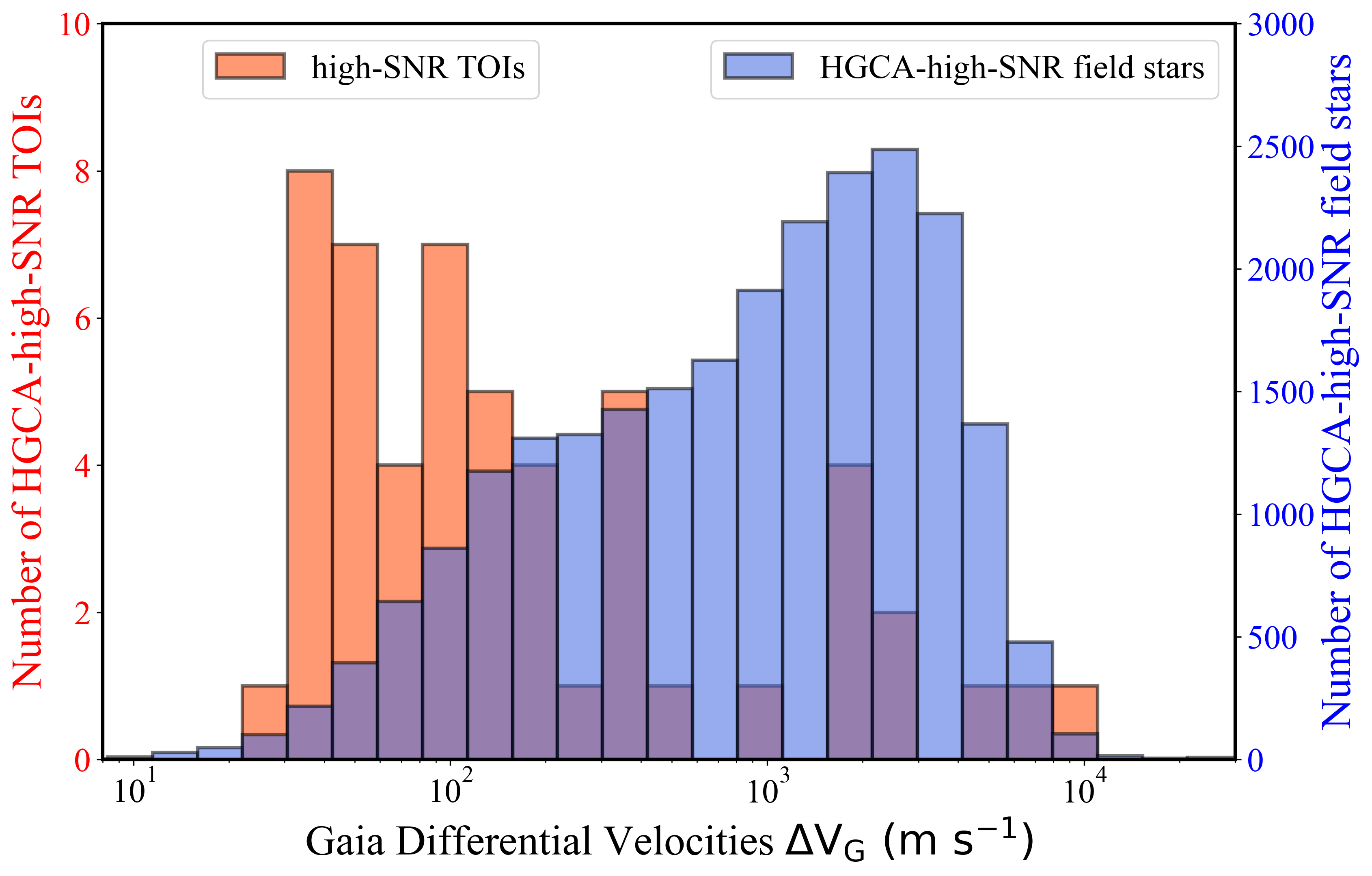}
    \caption{Gaia differential velocity distribution of TOIs with significant proper motion anomalies (red) compared to field stars (blue). The field star sample consists of all stars with significant proper motion anomalies from HGCA within 300 pc. The differential velocity are in log scale.  We exclude TOIs with FP/FA disposition.  }
    \label{fig:figuren}
\end{figure}

As detailed in section~\ref{sec:HGA}, the differential velocities can be approximately seen as the projected orbital velocities of the primary stars around the system barycenter, which increases with the companion masses and decreases with orbital distances. In this section, we compare the Gaia differential velocities of HGCA-high-SNR TOIs with field stars, which consists of all stars exhibiting significant proper motion anomalies from HGCA within 300 pc. Figure~\ref{fig:figuren} presents the results. The differential velocities of field stars exhibit a peak around 3000 $m\ s^{-1}$ and a broader bump centered at $\sim 200 m\ s^{-1}$ to the left.  The velocity magnitude at the peak is consistent with differential velocities caused by stellar companions with orbital periods from a few years to a few hundred years (sensitive range for Hipparcos-Gaia proper motion anomalies  method). For example, a solar mass star would have a differential velocity of around 4400 $m\ s^{-1}$ with a 0.5 $M_{\odot}$ companion at an orbital period of approximately 25 years, assuming a face-on orbit. As the orbital periods increase or companion masses decrease, the stellar companions' velocities produce a tail at lower velocities.  For example, a 0.01 $M_{\odot}$ companion would cause a velocity of around 100 $m\ s^{-1}$ for a solar mass star at 25 years period.

Compared to field stars, the distribution of HGCA-high-SNR TOIs displays a higher peak at velocities around 100 $m\ s^{-1}$ with a shortfall at high differential velocities. These distributions suggest that transiting planets are more likely to form in binaries when the companions have lower masses or are at wider separations. Our results are compatible with the previous studies that planets are less common in close binary systems compared to single systems or wide binaries \citep{Wang2015, Kraus2016, Ziegler2020, Ziegler2021, Hirsch2021, Moe2021}. In a recent study, \cite{Moe2021} finds that the occurrence rate of planets in binaries with $a<10$ AU is roughly $15\%$ of that in single systems, while wide binaries with $a>200$ AU have similar planet occurrence rates as single stars. The recent ALMA high-resolution surveys also find that disks in multiple systems are smaller, fainter, and less long-lived than those in singles \citep{Cox2017,Akeson2019,Manara2019,Zurlo2020,Zagaria2023}. These findings support the theory that close companions tidally truncate the circumstellar discs and reduce the reservoir of material available to assemble planetary embryos \citep{Paczynski1977,Rudak1981,Jang-Condell2015,Pichardo2005,Zagaria2023}.

\section{Orbit characterization of benchmark system: LTT 1445 ABC}\label{sec:OPSA}

In the section, we present the results of proof-of-concept system LTT 145 ABC, for which we characterize the three dimensional orbits of the companion pair BC around A with RVs, Hipparcos-Gaia astrometric acceleration and relative astrometry from AO imaging.

\subsection{Background}
LTT 1445 ABC (TOI-455) is the closest M dwarf triple known to harbor multiple planets at a distance of 6.86 pc \citep{Rossiter1955, Luyten1957,Luyten1980, winters2019,Winters2022}. The hierarchical system consists of a primary LTT 1445 A ($0.268\ R_{\odot}$, $0.257\ M_{\odot}$) orbited by a M dwarf pair BC at a separation of $\sim 7^{\arcsec}$ \citep{Dieterich2012, Rodriguez2015}. LTT 1445 A has two transiting super-Earths and one non-transiting planet: LTT 1445 Ab ($P_{b}=5.36$ days, $r_{b}=1.3\ R_{\oplus}$, $m_{b}=2.87\pm{0.25}\ M_{\oplus}$), LTT 1445 Ac ($P_{c}=3.12$ days, $r_{c}<1.15\ R_{\oplus}$, $m_{c}=1.54\pm{0.2}\ M_{\oplus}$), LTT 1445 Ad ($P_{b}=24.3$ days, $m_{d}=2.72\pm{0.25}\ M_{\oplus}$) \citep{winters2019,Winters2022, LTT1445Ad2022}. The BC subsystem is a visual binary pair with a separation of $\sim 1^{\arcsec}$. Using archival astrometry from Fourth Interferometric Catalog (FIC) and Differential Speckle Survey Instrument (DSSI), \cite{winters2019} found that LTT 1445 C orbits around B in an eccentric and edge-on orbit with a period of $\sim36$ years ($\rm{e_{C,B}=0.5\pm{0.11}}$, $\rm{i_{C,B}=89^{\circ}.64\pm{0.13}}$, $\rm{a_{C,B}=8.1\pm{0.5} \rm{AU}}$, $\rm{\Omega_{C}=137.63^{\circ}\pm{0.19}}$). However, it is unclear how the companion pair orbit around the primary. Therefore, we utilized the Hipparcos-Gaia proper motion anomalies, combined with primary RVs and relative astrometry to characterize the orbit of LTT1445 BC pair around the primary A. Furthermore, we constrain the mutual inclination between orbital plane of C around B and that of the subsystems around the primary. LTT 1445 A is targeted by James Webb Space Telescope Cycle 1 GO Program 2708 (PI Z. Berta Thompson) to investigate the presence of an atmosphere of the planet b. Our characterization of the companion pair's orbit provides context for the dynamical stability of the system.

\subsection{Orbit Fitting}
 We use 9 archival RVs of LTT1445 A taken with HARPS between 2004 and 2013 from \cite{Trifonov2020} and 136 published RVs from 2019 to 2021 taken with 5 high-precision spectrographs including the W. M. Keck Observatory echelle spectrograph HIRES, ESPRESSO, HARPS, MAROON-X and PFS from \cite{Winters2022}. We also include 5 RVs we newly collected between Sep. 2021 and Jan. 2023 using HIRES (see Table~\ref{tab:rvs}). The proper motion anomalies  of LTT 1445 A at Hipparcos and Gaia epoch are from HGCA. Finally, we adopt two published relative astrometric measurements taken at 2003 and 2010 from \cite{Dieterich2012} and \cite{Rodriguez2015} (see Table~\ref{tab:bc_obs}). We consider BC pair as one object and use the relative astrometry of mass center of BC subsystem to primary A.    

 \begin{deluxetable}{lccc}

\tablecaption{LTT 1445A RVs}\label{tab:rvs}
\tablehead{\colhead{Time} & \colhead{RV} & \colhead{$\sigma_{\mathrm{RV}}$} & \colhead{Inst} \\ 
\colhead{(BJD - 2450000)} & \colhead{(m/s)} & \colhead{(m/s)} & \colhead{} } 
\startdata
9482.1 & 1.69 & 1.37 & HIRES \\
9587.81 & -1.54 & 1.37 & HIRES \\
9824.13 & 2.33 & 1.29 &  HIRES \\
9832.03 & 3.5 & 1.46 & HIRES \\
9947.73 & 0.45 & 1.2 &  HIRES \\
\enddata
\tablecomments{Times are in BJD - 2450000.0. The RV uncertainties do not include RV jitter. We present 5 unpublished HIRES RVs in this table. All RV data utilized in the orbit fitting, including those sourced from the literature, are available in a machine-readable format. }

\end{deluxetable}

 We use open source package \textit{orvara} \citep{Brandt2021b}, which performs a parallel-tempering Markov Chain Monte Carlo (MCMC) fitting. In total, our analysis uses 15 free parameters. Two of them are the masses of the host star ($M_{\rm{A}}$) and combined mass of companion pair ($M_{\rm{BC}}$). Six orbital parameters define the orbit of companion pair, including semi-major axis (a), inclination (i), longitude of the ascending node ($\Omega$), mean longitude at a reference epoch ($t_{ref}$) of 2455197.5 JD ($\lambda_{ref}$), and  the eccentricity (e) and the argument of periastron ($\omega$) in the form of $e\sin \omega$ and $e\cos \omega$. We also included six parameters to fit the zero-point for RV data from different instruments. As there was a fiber exchange for HARPS in 2015, we use different RV zero points for HARPS RVs taken before and after 2015. The last parameter is the intrinsic jitter of RV data. We ignore the transiting planets because their RVs amplitudes are expected to be smaller than the stellar jitter. The proper motion anomalies  from inner planets are also nearly zero because signals cancel out when the orbital periods are much shorter than the observing window duration of \HIP and \Gaia. We use the primary mass and companion masses from \cite{winters2019} as priors in our fitting and bound the jitter between 0 to 10 $m\ s^{-1}$. The likelihood is calculated by comparing the measured separations, position angles, absolute astrometry, and radial velocities to those of a synthetic orbit and assuming Gaussian errors \citep{Brandt2021b}. 
 \begin{figure*}
    \centering
    \includegraphics[width=\linewidth]{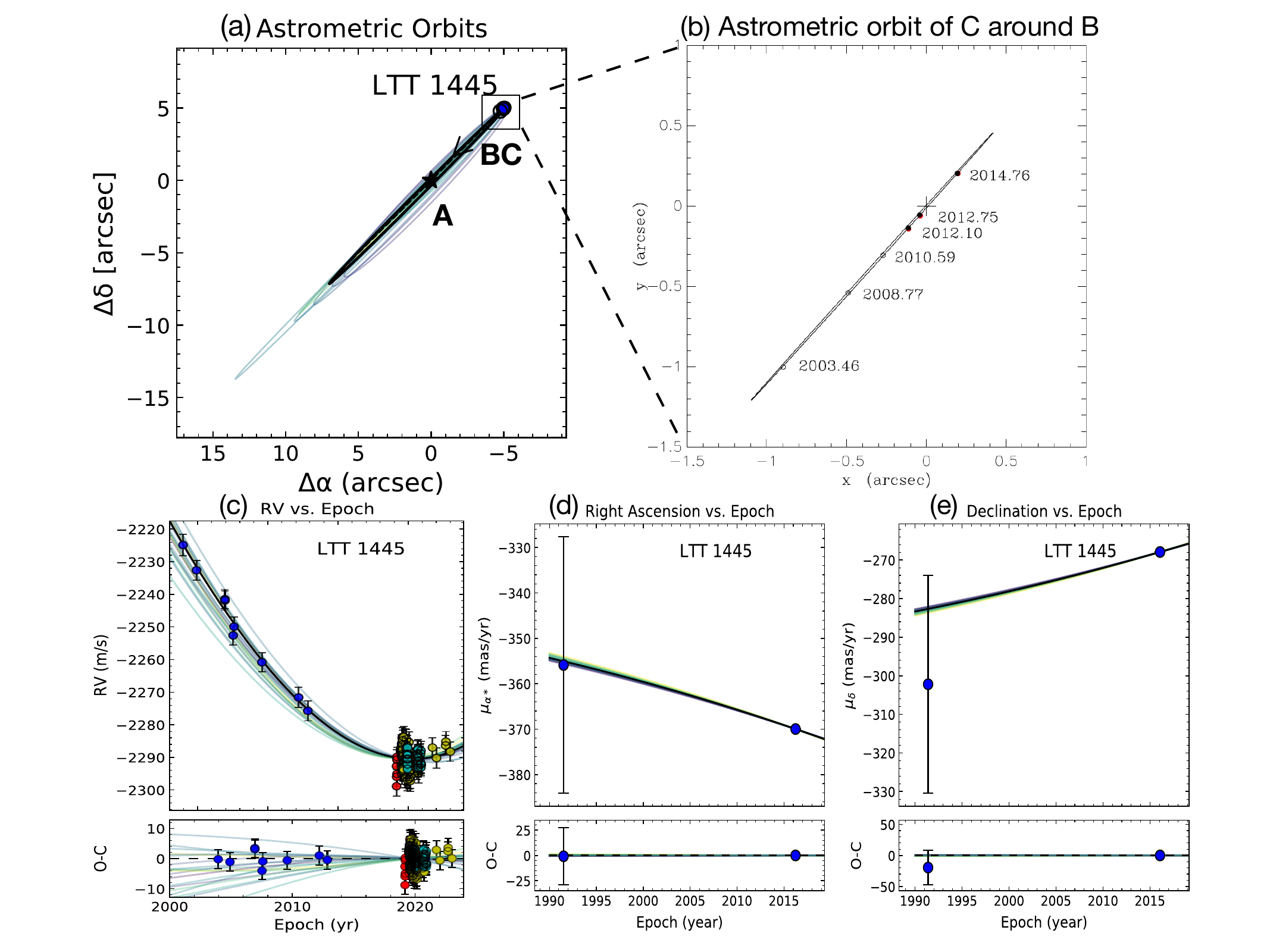}
    \caption{Orbit characterization of LTT 1445 BC mass center around A using RVs, relative astrometry and absolute astrometry from Hipparcos and Gaia.  (a): relative astrometry orbits of LTT 1445 BC pair around A. The blue filled circles are two observed relative astrometry used in our analysis. (c): Observed and fitted RVs of LTT 1445 from HARPS, HIRES, ESPRESSO, MAROON-X, PFS. (d)-(e): Observed and fitted Hipparcos and Gaia proper motion of LTT 1445 A in right ascension and declination. In all of above panels, the thicker black lines represent the best-fit orbit in the MCMC chain while the other 50 lines represent random draws from the chain; (b): relative astrometry orbits of LTT 1445 C  around B from \cite{winters2019}.}
    \label{fig:figuren2}
\end{figure*}
\begin{deluxetable}{llllcc}
\tabletypesize{\scriptsize}
\tablecaption{Relative Astrometry  used in the Orbit Characterization for LTT 1445BC around A}\label{tab:bc_obs}
\tablehead{ 
\colhead{Date} & &
\colhead{$\theta$} & \colhead{$\rho$} & 
\colhead{Instrument} & 
\colhead{Reference}\\
\colhead{} & &
\colhead{($^{\circ}$)} & 
\colhead{(${\prime \prime}$)} &
} 
\startdata 
2003.4620 & A-B&315.0 & 7.706  & HST/NICMOS & 1  \\
2003.4620 & B-C&138.1 & 1.344  & HST/NICMOS & 1  \\
2003.4620 & $\rm{A-BC}^{1}$&314.74 & 7.13  & HST/NICMOS & 3  \\
2010.594  & A-B&$313.79^{2}$ & 7.20  &  Lick/IRCAL&2  \\
2010.594  & B-C&$138.41$ & 0.41  &  Lick/IRCAL&2  \\
2010.594  & A-BC&$314.91$ & 7.02  &  Lick/IRCAL&3  \\
\enddata 
\tablenotetext{1}{Relative astrometry of the mass center of BC from A.}
\tablenotetext{2}{Quadrant ambiguous; the position angle here has been changed by 180 degrees relative to the original result.}
\tablerefs{
(1) \citet{Dieterich2012};
(2) \citet{Rodriguez2015};
(3) Derived in this work.}
\end{deluxetable}

 We use 100 walkers to sample our model and the chains converge after $2.5\times10^{5}$ steps. We discarded the first $30\%$ as burn-in portion. Figure \ref{fig:figuren2} shows the best-fit orbit (black lines) from our MCMC chains, including fitted astrometric orbit, RVs and Hipparcos-Gaia proper motions. The reduced $\chi^{2}$ of RVs indicates a good fit and accurate measurement errors,with values of 0.97. We obtain a dynamical mass of $\rm{M_{A}}={0.251}_{-0.010}^{+0.010}\ M_{\odot}$ for primary A and mass of $\rm{M_{BC}}={0.39}_{-0.009}^{0.009}\ M_{\odot}$ for BC subsystem, which agree with published values from \cite{Winters2022} within $1\sigma$. Our best-fit model shows that the subsystem BC orbits around primary A in an eccentric and edge-on orbit ($\rm{a_{BC,A}={58}_{+16}^{-20}}$ AU, $\rm{e_{BC,A}={0.375}_{-0.064}^{+0.037}}$, $\rm{i_{BC,A}={88^{\circ}.5}_{-1.4}^{+1.3}}$, $\rm{\Omega_{BC,A}={135.15^{\circ}}\pm{0.28}}$, see Table~\ref{tab:mcmc} for other parameters). We compute the mutual inclination $\Delta I$ between orbital plane of BC around A and that of C around B with their inclination ($\rm{i_{C,B}}$, $\rm{i_{BC,A}}$) and longitude of ascending node ($\rm{\Omega_{C,B}}$, $\rm{\Omega_{BC,A}}$):
\begin{equation}
    \rm{\cos \Delta i = \cos i_{C,B}\cos i_{BC,A} + \sin i_{C,B}\sin i_{BC,A} \cos (\Omega_{C,B}-\Omega_{BC,A}})
\end{equation}

We obtained the mutual inclination $\Delta i= 2^{\circ}.88\pm{0.63}$. Therefore, LTT 1445 ABC is a flat system where the subsystem BC orbits around A in nearly the same plane as their orbit around each other. Because the LTT 1445 A is a slow rotator ($P_{tot}\sim84$ days, \citealt{winters2019}), we are not able to measure the spin-orbit angle of two transiting planets b and c relative to the primary through Rossiter-MacLauglin effect. However, the probability of observing the two transiting planets and companion pair BC all have edge-on orbits is notably low, if we assume their orbits are independent. Specifically, if the $\cos (i)$ values of planet b, c, and companion pair BC are drawn randomly from a uniform distribution between 0 and 1, the probability of observing all these bodies to have inclinations within the range of $87^{\circ}$ to $90^{\circ}$ is only $0.014\%$. Therefore, it is highly likely that the transiting planet orbits are coplanar with the orbit of the BC companions. Meanwhile, the alignment of the non-transiting planet LTT 1445 Ad with the inner planets is subject to significant uncertainty. Located at a distance of 0.09 AU, the angle range for a transiting configuration is only $1.6^{\circ}$. A plausible scenario is that LTT 1445 Ad is aligned with the inner planets but is located outside the transiting configuration. A detailed dynamical study might yield interesting constraints on the possible orbits of planet d, but is outside the scope of this paper.

\begin{deluxetable*}{lcc}
\Large
\tablecaption{MCMC Orbital Posteriors for LTT 1445 \label{tab:mcmc}}
\setlength{\tabcolsep}{0.10in}
\tablewidth{0pt}
\tablehead{
\colhead{Parameter}              &
\colhead{Median $\pm$1$\sigma$} &
\colhead{Best-fit value}}
\startdata
\multicolumn{3}{c}{Stellar parameters} \\[3pt]
\cline{1-3}
\multicolumn{3}{c}{} \\[-5pt]
Host-star mass $\rm{M_{A}}\ (M_{\odot}) $                                      &  $0.25\pm{0.01}$                       &      0.25                                                          \\[3pt]
Companion pair mass $\rm{M_{BC}}\ (M_{\odot})$                                     & $0.39\pm{0.09}$                    &        0.39       \\[3pt]
\cline{1-3}
\multicolumn{3}{c}{Orbital parameters} \\[3pt]
\cline{1-3}
\multicolumn{3}{c}{} \\[-5pt]
Semi-major axis $a$ (AU)                                                     & ${58}_{-10}^{+16}$             &       63.45        \\[3pt]
Orbital period $P$ (yr)                                                     & ${549}_{-136}^{+243}$                   &    629.54                                                                           \\[3pt]
Inclination $i$ (deg)                                                   & ${88.5}_{-1.3}^{+1.3}$            &         88.82              \\[3pt]
$\sqrt{e}\sin{\omega}$                                                      & ${-0.568}_{-0.025}^{+0.036}$          &     -0.24        \\[3pt]
$\sqrt{e}\cos{\omega}$                                                      & ${-0.14}_{-0.24}^{+0.28}$        &           -0.58       \\[3pt]
Eccentricity $e$                                                            & ${0.375}_{-0.037}^{+0.084}$           &          0.4                                                              \\[3pt]
Mean longitude at $t_{\rm ref}=2455197.5$~JD, $\lambda_{\rm ref}$ (deg) & ${210.3}\pm{-2.1}$                  &           211.19         \\[3pt]
Longitide of the ascending node $\Omega$ (deg)                                 & $135.15\pm{0.28}$            &         135.10            \\[3pt]
Parallax (mas)                                                              & ${145.6923}_{-0.0040}^{+0.0040}$                   &    145.69       \\[3pt]
Argument of periastron $\omega$ (deg)                                   & ${256}_{-21}^{+28}$            &       247.52                                                                \\[3pt]
Time of periastron $T_0=t_{\rm ref}-P\frac{\lambda-\omega}{360 }$ (JD)& ${2480796}_{-5663}^{+5304}$          &     2478401.65                                                               \\[3pt]
\cline{1-3}
\multicolumn{3}{c}{} \\[-5pt]
\multicolumn{3}{c}{Other Parameters} \\[1pt]
\cline{1-3}
\multicolumn{3}{c}{} \\[-5pt]
RV jitter $\sigma$ (m\,s$^{-1}$)                                 & ${2.74}_{-0.18}^{+0.21}$  &     2.73      \\[3pt]
HARPS pre-2015 RV zero point (m\,s$^{-1}$)                                      & $3225.67^{+166}_{-177}$                       &          3162.27             \\[3pt]
HARPS post-2015 RV zero point (m\,s$^{-1}$)                                      & $-2188.44^{+166}_{-176}$                       &     -2250.74              \\[3pt]
EXPRESSO RV zero point (m\,s$^{-1}$)                                      & $3232.41^{+166}_{-177}$                       &          3169.01             \\[3pt]
MAROON-X RV zero point (m\,s$^{-1}$)                                      & $ -2226.99^{+166}_{-177}$                       &          -2290.44             \\[3pt]
HIRES RV zero point (m\,s$^{-1}$)                                      & $-2225.24^{+166}_{-177}$                       &          -2288.68             \\[3pt]
PSF RV zero point (m\,s$^{-1}$)                                      & $-2226.55^{+166}_{-177}$                       &          -2289.98           \\[3pt]
\enddata
\tablecomments{The $\chi^2$ of relative astrometry is 0.09 for separations and 0.06 for PAs, with 2 measurements for each. The $\chi^2$ of the Hipparcos and Gaia proper motion differences is 2.64 for four measurements. The $\chi^2$ of RV is 146.80 for 150 measurements.}
\end{deluxetable*}

\subsection{Implication for planet formation}
One piece that needs to be added to understand the effect of companions on planet formation is the inclination of the companion orbits. Inclination plays a vital role in the dynamic interaction between the companions and inner planets or the protoplanetary disks. For example, the Kozai-Lidov effect \citep{Kozai1962, Lidov1962} occurs when the mutual inclination between two objects is greater than $\sim 40^{\circ}$, causing the inner objects to be unstable. Previous studies have also found that an inclined outer companions may misaligned the orbit of inner planets \citep{Huber2013, Zhang2021}. Fortunately, the combination of Hipparcos and Gaia astrometry and RVs allow us to characterize the three-dimensional orbits of the companions. In this work, we present the results of triple system LTT 1445, in which LTT 1445 BC orbit around the primary A with a semi-major axis of $\sim 58 $AU.  The LTT 1445 system bears a remarkable resemblance to the Kepler-444 triple system, where Kepler-444 BC orbits around A in an edge-on orbit and is likely to be aligned with the orbit of five transiting planets around the primary A \citep{Dupuy2016,Zhangzj2022}. LTT 1445 and Kepler-444 both agree with the statistical results reported by \cite{Dupuy2022}, which concludes that low mutual inclinations between planets and companions are required to explain the observed orbital arcs in 45 binary systems containing \textit{Kepler} planet candidates. One may ponder whether the coplanarity of the systems relates to the planet's formation in a dynamically hostile environment.

\cite{Zanazzi2018} investigated the evolution of disk inclinations in binary systems and found that effective realignment between the circumstellar disks around the primary star and companions tends to occur when the companions are closer than 200 AU. Considering that LTT1445 BC has a semi-major axis of $\sim 58$ AU, it is plausible that the companions were initially misaligned with the primary disk but later underwent realignment during their evolution. Another close binary HIP94235 ($a \sim 50 AU$) is consistent with the possibility in which the primary hosts a transiting mini-Neptune. The companion HIP94235B exhibits an inclination of around  $68^{\circ}$ \citep{zhou2022}, which suggests a minimum misalignment of $22^{\circ}$ between the companions and the transiting planet. Given that HIP94235 is part of a young comoving group ($\sim 120$ Myr, \citealt{zhou2022}), it is possible that the realignment between the companion and the disk is still ongoing. Alternatively, it is also possible that the LTT 1445 BC were initially aligned with the primary disk from the onset. In this case, the fully con-planar configuration and close separation of LTT 1445 triples are consistent with the disk fragmentation scenario that gravitational
instability in a shearing disk might produce multiples stars \citep{Adams1989,MK2018,Offner2022}. In either possibility, although the companions likely truncated the primary's circumstellar disk, there was still enough disk material remaining to form multiple planets.  In short, constraining the mutual inclination between planets and stellar companions in more systems is needed to understand the mechanisms behind planet formation in close binary systems.

\section{Conclusions}\label{sec:conc}
We have presented a study of transiting planets with systems that show significant Gaia-Hipparcos accelerations. Our conclusions are as follows:

\begin{enumerate}
    \item We presented a catalog of 66 transiting planet hosts (58 TOIs, 4 KOIs, and 4 K2 planet candidates) within a 300 pc volume limit with significant \HIP and \Gaia proper motion anomalies  through cross-matching the TOI/KOI/K2 catalogs with HGCA. The parameters of these targets are presented in Table~\ref{tab:table1}. Among these targets, 33 have directly imaged stellar companions, either from published papers or the ExFOP website.  

    \item For transiting planet candidates identified by \textit{TESS}, we evaluated the reliability of the transits based on the radial velocities obtained with Keck/HIRES and the TESS Follow-Up Observing Program. We found that TOIs with high proper motion anomalies  have nearly four times more eclipsing binary classifications than TOIs with insignificant proper motion anomaliess. The excess of EBs in HGCA-high-SNR TOIs might be from the contamination of triple systems. 

    \item  We translated the proper motion anomalies  into differential velocities between the epochs of Hipparcos and Gaia, expressed in units of $m s^{-1}$. We observe a tentative inverse correlation between transiting planet orbital periods and \Gaia differential velocities, with short planet periods occurring preferentially with more massive and closer companions. Additionally, our findings suggest a possible trend of shorter planet periods in binaries, although this could be an artifact of the \TESS observation bias. If the trend is genuine, it supports the theory that planets in binaries form in smaller protoplanetary disks truncated by their companions. 

    \item We observe that HGCA-high-SNR TOIs exhibit lower differential velocities than  the entire population of significant proper motion anomalies stars within 300 pc in HGCA catalog. This comparison indicates that planets are more likely to persist in systems with low-mass companions or wider stellar companions.

    \item We determined the three-dimensional orbit of the companion pair BC around the primary star A in the triple system LTT 1445, which also hosts two transiting planets. Our analysis indicates that LTT 1445 is a flat system, with the orbital plane of BC around A being almost coplanar with the orbital plane of the outer planet c around B ($\Delta i\sim 2.88^{\circ}$). This coplanarity may account for the survival of multiple planets in an otherwise dynamically challenging environment. 
\end{enumerate}

Future observations will provide opportunities to confirm potential companions in systems with no reported companions. Our next paper in the series will feature AO images and astrometric measurements. We will also constrain the companion mass and separation for low-mass companions below the detection limit. Additionally, we will identify the host stars based on transit duration and stellar density and recalculate planet radii by estimating the contrast between the two stars.

\acknowledgments
 
J.Z. would like to thank Jerry Xuan, Pierre Kervella1, and Michael Liu for helpful discussions. D.H. acknowledges support from the Alfred P. Sloan Foundation, the National Aeronautics and Space Administration (80NSSC21K0652), and the Australian Research Council (FT200100871).  L.M.W. acknowledges support from the NASA-Keck Key Strategic Mission Support program (grant no. 80NSSC19K1475) and the NASA Exoplanet Research Program (grant no. 80NSSC23K0269).
This work was supported by a NASA Keck PI Data Award, administered by the NASA Exoplanet Science Institute. Data presented herein were obtained at the W. M. Keck Observatory from telescope time allocated to the National Aeronautics and Space Administration through the agency’s scientific partnership with the California Institute of Technology and the University of California. The Observatory was made possible by the generous financial support of the W. M. Keck Foundation.

Some of the observations in this paper made use of the High-Resolution Imaging instruments ‘Alopeke and Zorro and were obtained under Gemini LLP Proposal Number: GN/S-2021A-LP-105. ‘Alopeke and Zorro were funded by the NASA Exoplanet Exploration Program and built at the NASA Ames Research Center by Steve B. Howell, Nic Scott, Elliott P. Horch, and Emmett Quigley. Alopeke was mounted on the Gemini North telescope of the international Gemini Observatory, a program of NSF's OIR Lab, which is managed by the Association of Universities for Research in Astronomy (AURA) under a cooperative agreement with the National Science Foundation. on behalf of the Gemini partnership: the National Science Foundation (United States), National Research Council (Canada), Agencia Nacional de Investigación y Desarrollo (Chile), Ministerio de Ciencia, Tecnología e Innovación (Argentina), Ministério da Ciência, Tecnologia, Inovações e Comunicações (Brazil), and Korea Astronomy and Space Science Institute (Republic of Korea). 

Funding for the TESS mission is provided by NASA's Science Mission Directorate. This research has made use of the Exoplanet Follow-up Observation Program website, which is operated by the California Institute of Technology, under contract with the National Aeronautics and Space Administration under the Exoplanet Exploration Program.

This research has made use of the NASA Exoplanet Archive and ExoFOP, which are operated by the California Institute of Technology, under contract with the National Aeronautics and Space Administration under the Exoplanet Exploration Program.

The authors wish to recognize and acknowledge the very significant cultural role and reverence that the summit of Mauna Kea has always had within the indigenous Hawaiian community. We are most fortunate to have the opportunity to conduct observations from this mountain.

\vspace{5mm}
\facilities{\kepler, \TESS, \Gaia, \HIP, Keck:10m}


\software{\textit{orvara} \citep{Brandt2021b},  
          }



\appendix
\counterwithin{figure}{section}
\section{Notes on false positives in the target sample}\label{sec:ap1}

\begin{itemize}
    \item HIP33681 (TOI-510): No obvious NEBs from SG1 but cannot yet clear very close neighbor.
    \item HIP15053 (TOI-394): TFOP SG1 identified different odd-even depth in TESS QLP light curve, which indicates that TOI394.01 is an EB. The RVs at two quadrature time collected using Keck/HIRES from the primary don't show a RV difference consistent with a stellar mass object. So it is likely that the primary is orbited by a close EB companion and the transit is from the companion pair instead of the primary.
    \item HIP21952(TOI-271): TFOP SG1 finds no NEBs.
    \item HIP98516 (TOI-1124): TFOP SG1 detected event on target. But TFOP SG1 also detected strong chromaticity of the transit depth and odd-even depth in TESS light curve, which indicates TOI-1124.01 is a blend. 
    \item HIP28122 (TOI-896): TFOP SG1 evaluated it as false alarm because of marginal signal. No transiting events are detected in sector 33, therefore TOI-896.01 retired as false alarm. 
    \item HIP83168 (TOI-1418): TFOP SG1 identified multi-sector data seem more consistent with stellar variability.
    \item HIP58234 (TOI-680): SG1 clears the field of NEBs, and detects a 1ppt event arriving a little early. Additional transits show chromaticity, and HIRES RVs show no convincing variation phased to the ephemeris. This is likely a blend or a planet around the compaion. 
    \item HIP210000 (TOI-953): TFOP SG1 notes that WASP follow-up RVs show this to be an EB.
    \item HIP22084 (TOI-4314): TOI4314.01 has a long period of ~73 days. TFOP SG1 note that the transits have low SNRs and possibly came from stellar valibility. 
    \item HIP79105 (TOI-2118): TFOP SG1 finds NEB 43" E.
    \item HIP55069 (TOI-1204): No obvious NEB in the SG1 sheet.
    \item HIP70833 (TOI-1946): TOI1946.01 retired as TFOP FP/NEB.
    \item HIP27844 (TOI-1665): TFOP SG1 finds NEB 32" E.
    \item HIP67650 (TOI-1837): TFOP SG1 notes that TRES+FIES RVs reveal 33 km/s variation consistent with
 stellar companion.
    \item HIP45621 (TOI-2666): TOI2666.01 is a single transit. Keck/HIRES spectra show that the star is a spectroscopic binary(SB). The companion is 18.5$\%$ the brightness of the secondary and separated by 35 km/s.
    \item HIP13754 (TOI-179): TFOP SG1 detects event on target
    \item HIP108162 (TOI-1099): TFOP SG1 detects event on target.
    \item HIP57386 (TOI-5521): There are no SG1 results available for this system. However, two TRES observations indicate a velocity offset of 29.5 km/s that is out of phase with the photometric ephemeris. This strongly suggests the presence of a stellar companion in the system, although it cannot be responsible for the shallow transits observed. This is consistent with the WDS catalog, which presents a companion at $0.7^{\arcsec}$ to TOI-5521.
    \item HIP93711 (TOI-2299): Possible on target. SG1 detected the transit, but it is unclear whether it originated from the target or a neighbor at 3.6 arcsec to the west. In addition, spoc-s14-s60 detects two TCEs at 214 and 246 days instead of a single TCE at 165 days. But they may not be reliable.
    \item HIP57990 (TOI-635): TFOP SG1 notes that TOI-634, TOI-635, TOI-638 all have similar ephemeris. Possible false alarm.
    \item HIP65205 (TOI-1831): It's a large star with a close companion at 0.66 arcsecs. The transit shows a slight odd-even transit depth difference, which could possibly come from an EB. SG1 clears the field of NEBs. But TRES observation does not find a large RV variation, so it retired as an APC/EB?/CRV in SG1. However, as there is a very close companion at 0.66 arcsec from the star, it is also possible that the transit event is occurring at the companion, which could explain why large RV variations were not seen in the primary star. We plan to investigate this system further in our project.
    \item HIP24718 (TOI-128): TFOP SG1 finds no NEBs beyond $2^{\arcsec}$ but they cannot rule out the close companion at $\sim2^{\arcsec}$. If the signal originates from either the primary star or companion, it could still be a planet.
    \item HIP82032 (TOI-909): TFOP SG1 detects an NEB on a nearby, $\Delta\ T$=6.3 star, TIC 1310226289. This is consistent with the SPOC centroids.
    \item HIP41849 (TOI-575): TFOP SG1 notes additional TESS data reveal this to be an EB, with primaries and secondaries both visible. Probably on a 0.6" companion seen in high-resolution imaging. 
    \item HIP74685 (TOI-2017): TFOP SG1 detects event on target. But TOI2017.01 is an F+M EB with an orbital solution from TRES and the CfA Digital Speedometers.
    \item HIP118045 (TOI-222): TFOP SG1 identified as Spectroscopic Eclipsing Binary(SEB2). 
    \item HIP40694 (TOI-522): TFOP SG1 clears the field of NEBs. HIP40694 is a rapid rotator, with a vsini of ~151 km/s.  
    \item HIP54491 (TOI-1799): TFOP SG1 clears the field of NEBs. 
    \item HIP78301 (TOI-906): TFOP SG1 detects event in aperture also containing 1", $\Delta\ T$=2.9 companion.
    \item HIP94780 (K06139): Kepler Eclipsing Binary Catalog v2 \citep{Slawson2011} marks K06139.01 as an EB.
    \item HIP96501 (K01924): KOI 1924.01 is a false positive due to an eclipsing binary 77 arcseconds away.
    \item  HIP93954 (K06364): Kepler Eclipsing Binary Catalog v2 \citep{Slawson2011} marks K06364.01 as an EB.
    \item HIP80474 (EPIC204165788): \cite{Barros2016} marks it as eclipsing binary.
    \item HIP54766 (EPIC201488365): \cite{Armstrong2015} marks it as eclipsing binary.
    \item HIP41431 (EPIC212096658): \cite{Kruse2019} marks it as eclipsing binary.
    \item HIP78977 (EPIC204506777): \cite{Rizzuto2017} marks it as eclipsing binary.
\end{itemize}

\newpage

\section{LTT 1445 MCMC fitting plots}\label{sec:ap3}
\begin{figure*}[h]
    \centering
    \includegraphics[width=0.5\linewidth]{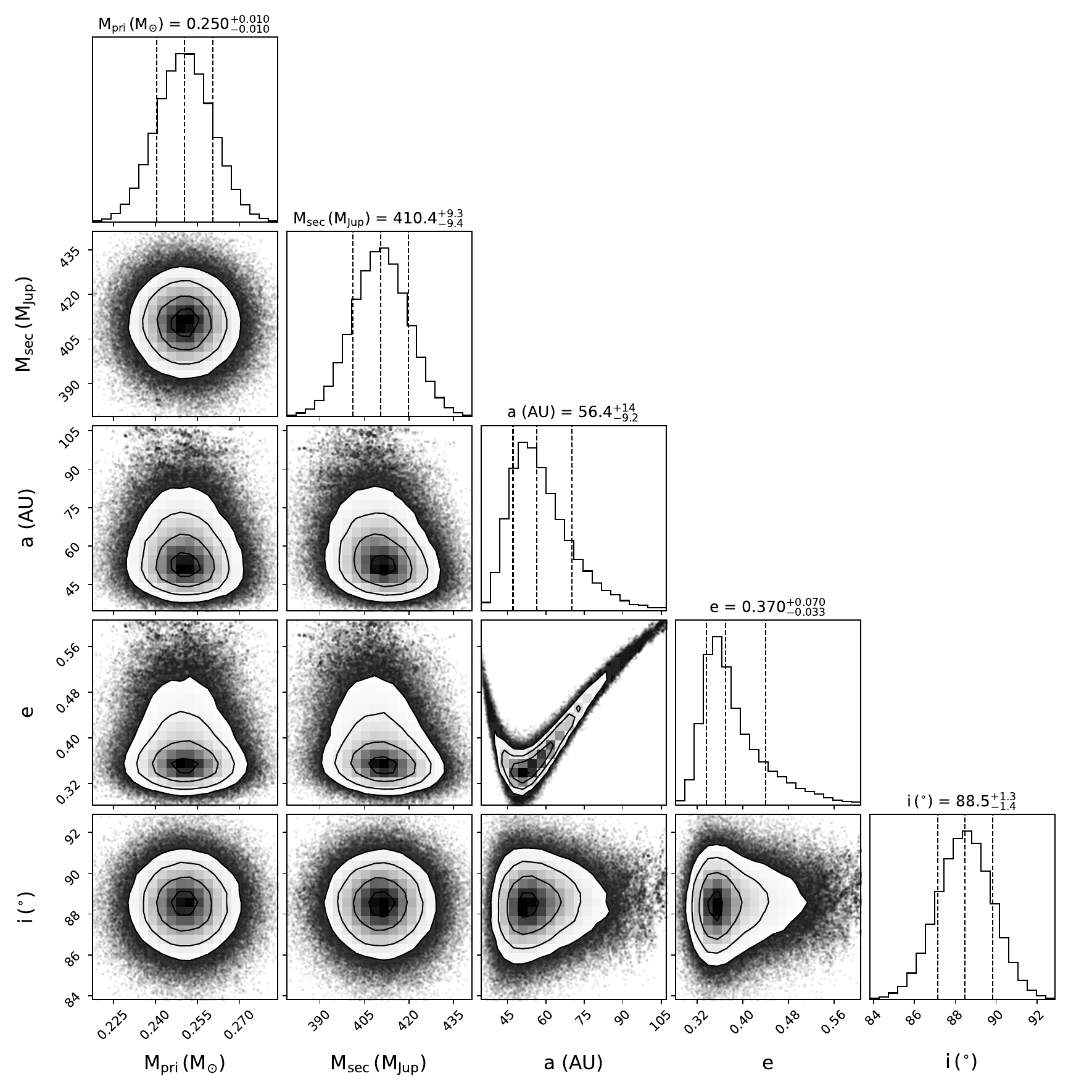}
    \caption{Joint posterior distributions for selected orbital parameters of LTT 1445 BC. These are the host star's mass (Mpri), the companion mass (Msec), the semi-major axis a, the orbital eccentricity e, and the orbital inclination i. The values and histogram distributions of the posteriors of selected parameters are shown, along with 1 $\sigma$ uncertainties.}
    \label{fig:figurea1}
\end{figure*}

\begin{figure*}[!b]
    \centering
    \includegraphics[width=0.6\linewidth]{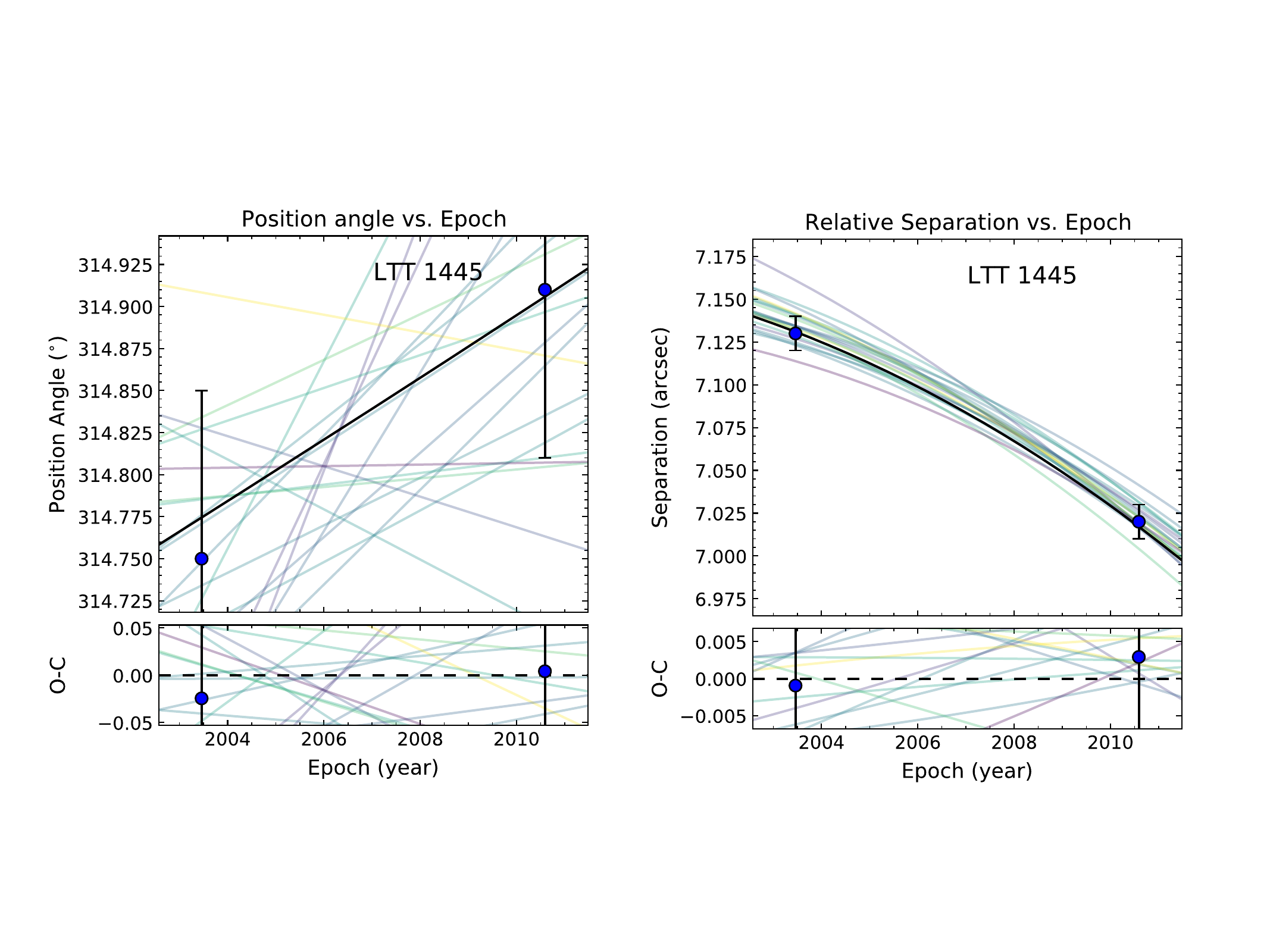}
    \caption{Observed and fitted absolute astrometry for the LTT 1445 system. The two panels show the  position angle and relative separation of LTT 1445 BC. The thicker black lines represent the best-fit orbit in the MCMC chain while the other 20 lines represent random draws from the chain.}
    \label{fig:figurea3}
\end{figure*}

\newpage
\section{TOIs with insignificant proper motion anomalies ($\rm{SNR_{G}}<3$) within 300 pc}\label{sec:toilow}
\begin{center}
\begin{longtable*}{lcccccc}
\caption{TOIs with low-SNR proper motion anomalies}\label{tab:tablea1}\\
\hline
Name & HIP & $P_{pl}$ & Exoplanet Archive & $\Delta v_{G}$ &  $\rm{SNR_{G}}$ & distance \\
 &  Number &  days&  Disposition &  $m\ s^{-1}$ &   & pc   \\
\hline
\hline
\endfirsthead
\hline
Name & HIP & $P_{pl}$ & Exoplanet Archive & $\Delta v_{G}$ &  $\rm{SNR_{G}}$ & distance \\
 &  Number &  days&  $\rm{Disposition^{a}}$ &  $m\ s^{-1}$ &   & pc   \\
\hline
\hline
\endhead
\hline
\endfoot
TOI 1011 & 36964 & 2.47 & PC & 3.81 & 0.47 & 52.38 \\
TOI 1014 & 33170 & 1.41 & PC & 65.65 & 1.27 & 219.9 \\
TOI 1025 & 46594 & 9.68 & PC & 19.66 & 0.34 & 130.12 \\
TOI 1028 & 51271 & 1.02 & FP & 4.33 & 1.18 & 23.74 \\
TOI 1029 & 27969 & 36.22 & FP & 22.25 & 2.09 & 68.69 \\
TOI 1053 & 95348 & 5.74 & FP & 76.28 & 0.82 & 266.02 \\
TOI 1054 & 99212 & 15.51 & CP & 14.43 & 1.02 & 89.05 \\
TOI 1055 & 96160 & 17.47 & CP & 6.81 & 0.74 & 56.79 \\
TOI 1097 & 61723 & 9.19 & CP & 18.23 & 1.45 & 79.56 \\
TOI 1097 & 61723 & 13.9 & CP & 18.23 & 1.45 & 79.56 \\
TOI 1098 & 62662 & 10.18 & CP & 15.4 & 0.89 & 105.14 \\
TOI 1104 & 91434 & 341.28 & PC & 103.48 & 1.64 & 68.51 \\
TOI 1127 & 95774 & 2.32 & FP & 35.04 & 0.8 & 174.91 \\
TOI 1135 & 62908 & 8.03 & PC & 13.44 & 0.71 & 114.18 \\
TOI 1148 & 52796 & 5.55 & KP & 9.68 & 0.59 & 96.88 \\
TOI 1150 & 101252 & 1.48 & KP & 16.27 & 0.42 & 207.22 \\
TOI 119 & 31609 & 5.54 & PC & 9.89 & 0.71 & 66.76 \\
TOI 119 & 31609 & 10.69 & PC & 9.89 & 0.71 & 66.76 \\
TOI 120 & 1419 & 11.54 & CP & 19.46 & 1.53 & 80.04 \\
TOI 1203 & 54779 & 25.52 & CP & 2.93 & 0.28 & 64.96 \\
TOI 1207 & 37931 & 2.63 & PC & 29.16 & 1.16 & 113.76 \\
TOI 1233 & 60689 & 14.18 & CP & 12.12 & 1.37 & 64.57 \\
TOI 1233 & 60689 & 19.59 & CP & 12.12 & 1.37 & 64.57 \\
TOI 1233 & 60689 & 6.2 & CP & 12.12 & 1.37 & 64.57 \\
TOI 1233 & 60689 & 3.8 & CP & 12.12 & 1.37 & 64.57 \\
TOI 1247 & 74326 & 15.92 & PC & 21.43 & 1.76 & 73.39 \\
TOI 1250 & 88071 & 1.44 & FA & 39.3 & 1.44 & 66.44 \\
TOI 1255 & 97166 & 10.29 & CP & 11.41 & 0.97 & 66.04 \\
TOI 129 & 65 & 0.98 & CP & 42.74 & 2.43 & 61.87 \\
TOI 130 & 3911 & 14.34 & CP & 12.34 & 1.13 & 57.41 \\
TOI 134 & 115211 & 1.4 & CP & 16.69 & 1.68 & 25.18 \\
TOI 1354 & 102712 & 1.43 & FP & 323.06 & 1.94 & 248.45 \\
TOI 1355 & 109028 & 2.17 & PC & 34.12 & 0.82 & 250.08 \\
TOI 139 & 110692 & 11.07 & PC & 31.29 & 2.52 & 42.42 \\
TOI 1407 & 110758 & 9.98 & KP & 14.42 & 0.77 & 80.5 \\
TOI 1411 & 76042 & 1.45 & CP & 17.1 & 2.32 & 32.46 \\
TOI 1415 & 71409 & 14.42 & PC & 35.65 & 2.03 & 111.98 \\
TOI 1416 & 70705 & 1.07 & PC & 9.71 & 0.85 & 55.04 \\
TOI 1430 & 98668 & 7.43 & PC & 3.88 & 0.57 & 41.24 \\
TOI 1431 & 104051 & 2.65 & CP & 47.14 & 1.87 & 149.59 \\
TOI 1434 & 57350 & 29.9 & PC & 9.64 & 1.52 & 37.81 \\
TOI 1440 & 92848 & 15.52 & PC & 161.27 & 1.94 & 236.48 \\
TOI 1440 & 92848 & 4.63 & PC & 161.27 & 1.94 & 236.48 \\
TOI 1462 & 85268 & 2.18 & CP & 6.89 & 1.46 & 27.01 \\
TOI 1471 & 9618 & 20.77 & PC & 23.86 & 1.66 & 67.29 \\
TOI 1471 & 9618 & 683.33 & PC & 23.86 & 1.66 & 67.29 \\
TOI 1475 & 117382 & 8.5 & PC & 32.14 & 0.75 & 289.64 \\
TOI 1487 & 83359 & 23.29 & FA & 14.66 & 0.62 & 109.73 \\
TOI 1488 & 85444 & 0.47 & FA & 63.85 & 2.34 & 159.7 \\
TOI 1514 & 115710 & 1.37 & FP & 123.97 & 2.77 & 239.68 \\
TOI 1573 & 13192 & 21.22 & KP & 30.16 & 2.56 & 77.43 \\
TOI 1599 & 11397 & 1.22 & KP & 43.04 & 1.34 & 121.6 \\
TOI 1608 & 15767 & 2.47 & APC & 3.96 & 0.15 & 101.29 \\
TOI 1611 & 107038 & 16.2 & CP & 5.13 & 1.1 & 28.28 \\
TOI 1652 & 61278 & 0.67 & PC & 23.65 & 1.27 & 129.3 \\
TOI 1682 & 24323 & 2.73 & KP & 14.76 & 0.72 & 135.52 \\
TOI 1683 & 20528 & 3.06 & PC & 43.52 & 2.16 & 50.94 \\
TOI 1689 & 84062 & 9.12 & PC & 48.89 & 0.4 & 28.02 \\
TOI 1718 & 36272 & 5.59 & PC & 8.34 & 0.82 & 52.06 \\
TOI 1726 & 38228 & 7.11 & CP & 9.75 & 2.39 & 22.38 \\
TOI 1726 & 38228 & 20.54 & CP & 9.75 & 2.39 & 22.38 \\
TOI 173 & 10389 & 29.75 & PC & 27.7 & 1.05 & 150.07 \\
TOI 174 & 17264 & 17.67 & CP & 1.14 & 0.14 & 39.09 \\
TOI 174 & 17264 & 29.8 & CP & 1.14 & 0.14 & 39.09 \\
TOI 174 & 17264 & 12.16 & CP & 1.14 & 0.14 & 39.09 \\
TOI 174 & 17264 & 3.98 & CP & 1.14 & 0.14 & 39.09 \\
TOI 174 & 17264 & 7.91 & CP & 1.14 & 0.14 & 39.09 \\
TOI 177 & 6365 & 2.85 & CP & 4.15 & 0.44 & 22.45 \\
TOI 1773 & 43587 & 0.74 & KP & 4.35 & 1.18 & 12.59 \\
TOI 1774 & 48443 & 16.71 & CP & 10.68 & 1.02 & 53.84 \\
TOI 1776 & 53688 & 2.8 & PC & 8.38 & 0.87 & 44.75 \\
TOI 1777 & 49576 & 14.65 & PC & 15.73 & 0.77 & 80.14 \\
TOI 1778 & 44746 & 6.52 & PC & 41.16 & 1.64 & 99.31 \\
TOI 1793 & 53719 & 55.09 & CP & 10.92 & 1.74 & 36.97 \\
TOI 180 & 4460 & 0.84 & APC & 159.24 & 2.21 & 259.7 \\
TOI 1801 & 57099 & 10.64 & PC & 16.67 & 1.12 & 30.89 \\
TOI 1807 & 65469 & 0.55 & CP & 14.14 & 1.66 & 42.59 \\
TOI 1821 & 54906 & 9.49 & KP & 1.72 & 0.42 & 21.56 \\
TOI 1827 & 62452 & 1.47 & CP & 4.12 & 1.06 & 8.08 \\
TOI 1830 & 73765 & 9.78 & FP & 19.35 & 2.74 & 50.31 \\
TOI 185 & 7562 & 0.94 & KP & 55.25 & 2.63 & 122.79 \\
TOI 186 & 16069 & 35.61 & CP & 2.26 & 0.51 & 16.33 \\
TOI 186 & 16069 & 7.79 & CP & 2.26 & 0.51 & 16.33 \\
TOI 1860 & 73869 & 1.07 & CP & 5.4 & 0.74 & 45.74 \\
TOI 1898 & 47288 & 45.52 & PC & 43.35 & 2.52 & 80.13 \\
TOI 193 & 117883 & 0.79 & CP & 12.65 & 0.8 & 81.05 \\
TOI 196 & 4548 & 1.16 & PC & 100.38 & 2.26 & 290.64 \\
TOI 1969 & 72490 & 2.69 & FP & 45.06 & 2.28 & 106.82 \\
TOI 197 & 116158 & 14.28 & CP & 6.35 & 0.27 & 95.43 \\
TOI 198 & 738 & 10.22 & PC & 11.5 & 1.5 & 23.78 \\
TOI 200 & 116748 & 8.14 & CP & 24.44 & 2.78 & 44.18 \\
TOI 2009 & 5286 & nan & PC & 13.94 & 2.42 & 20.54 \\
TOI 2018 & 74981 & 7.44 & PC & 4.7 & 0.73 & 28.04 \\
TOI 2020 & 80076 & 5.63 & KP & 12.17 & 0.63 & 128.52 \\
TOI 2024 & 80838 & 2.88 & KP & 9.31 & 0.77 & 76.22 \\
TOI 2056 & 848 & 10.22 & PC & 15.34 & 1.1 & 92.74 \\
TOI 2069 & 80243 & 5.92 & PC & 10.33 & 1.74 & 38.61 \\
TOI 2082 & 78892 & 30.2 & PC & 23.37 & 2.35 & 63.74 \\
TOI 2091 & 86067 & 177.22 & PC & 10.72 & 0.9 & 70.39 \\
TOI 2105 & 58868 & 15.92 & PC & 33.36 & 1.84 & 72.89 \\
TOI 2111 & 84840 & 1.27 & FP & 35.33 & 2.74 & 83.16 \\
TOI 2112 & 113195 & 14.01 & PC & 9.16 & 0.69 & 86.17 \\
TOI 2112 & 113195 & 155.82 & PC & 9.16 & 0.69 & 86.17 \\
TOI 2115 & 6105 & 3.69 & FP & 13.0 & 0.27 & 215.64 \\
TOI 2128 & 83827 & 16.34 & PC & 4.49 & 0.93 & 36.67 \\
TOI 214 & 31692 & 18.55 & PC & 13.43 & 1.11 & 38.94 \\
TOI 214 & 31692 & 9.7 & PC & 13.43 & 1.11 & 38.94 \\
TOI 2145 & 86040 & 10.26 & CP & 49.45 & 1.36 & 226.24 \\
TOI 2194 & 98130 & 15.34 & PC & 1.97 & 0.42 & 19.55 \\
TOI 2211 & 101503 & 3.09 & PC & 3.86 & 0.19 & 70.45 \\
TOI 2221 & 102409 & nan & CP & 2.48 & 1.05 & 9.71 \\
TOI 2259 & 79876 & 16.59 & PC & 39.79 & 2.09 & 121.94 \\
TOI 2270 & 79823 & nan & PC & 4.37 & 0.29 & 94.23 \\
TOI 2301 & 74576 & 6.05 & PC & 23.7 & 1.28 & 119.01 \\
TOI 2431 & 11707 & 0.22 & PC & 15.13 & 1.1 & 36.01 \\
TOI 2443 & 12493 & 15.67 & PC & 5.98 & 0.9 & 23.91 \\
TOI 245 & 113831 & 8.77 & PC & 80.43 & 2.27 & 125.06 \\
TOI 2474 & 24830 & 4.28 & PC & 40.22 & 1.21 & 132.43 \\
TOI 248 & 10779 & 5.99 & PC & 7.95 & 0.68 & 76.09 \\
TOI 253 & 4468 & 3.51 & PC & 3.02 & 0.32 & 30.91 \\
TOI 2540 & 25775 & 12.72 & PC & 4.15 & 0.51 & 19.16 \\
TOI 2540 & 25775 & 22.08 & APC & 4.15 & 0.51 & 19.16 \\
TOI 257 & 14710 & 18.39 & CP & 21.14 & 1.85 & 76.86 \\
TOI 260 & 1532 & 13.48 & PC & 4.27 & 0.88 & 20.21 \\
TOI 261 & 4739 & 3.36 & PC & 28.81 & 0.9 & 113.64 \\
TOI 261 & 4739 & 13.04 & CP & 28.81 & 0.9 & 113.64 \\
TOI 262 & 10117 & 11.15 & CP & 3.72 & 0.43 & 44.14 \\
TOI 266 & 8152 & 10.75 & PC & 4.92 & 0.23 & 101.69 \\
TOI 266 & 8152 & 6.19 & PC & 4.92 & 0.23 & 101.69 \\
TOI 282 & 20295 & 56.01 & CP & 16.46 & 0.74 & 140.24 \\
TOI 282 & 20295 & 31.32 & FA & 16.46 & 0.74 & 140.24 \\
TOI 282 & 20295 & 84.26 & CP & 16.46 & 0.74 & 140.24 \\
TOI 282 & 20295 & 22.89 & CP & 16.46 & 0.74 & 140.24 \\
TOI 381 & 7060 & 4.9 & FP & 14.48 & 1.42 & 75.3 \\
TOI 387 & 16212 & 4.16 & FP & 37.28 & 0.56 & 218.8 \\
TOI 389 & 36612 & 13.46 & FP & 20.69 & 1.16 & 105.0 \\
TOI 396 & 13363 & 5.97 & CP & 6.67 & 1.01 & 31.7 \\
TOI 396 & 13363 & 3.59 & CP & 6.67 & 1.01 & 31.7 \\
TOI 396 & 13363 & 11.23 & CP & 6.67 & 1.01 & 31.7 \\
TOI 409 & 33392 & 6.8 & FP & 16.24 & 1.46 & 53.27 \\
TOI 411 & 17047 & 9.57 & CP & 6.89 & 0.6 & 62.9 \\
TOI 411 & 17047 & 4.04 & CP & 6.89 & 0.6 & 62.9 \\
TOI 4186 & 105697 & 12.76 & PC & 42.92 & 2.22 & 67.61 \\
TOI 4187 & 14982 & 30.88 & PC & 45.53 & 1.99 & 154.81 \\
TOI 4189 & 25359 & 46.96 & PC & 6.95 & 0.63 & 69.31 \\
TOI 419 & 33390 & 0.4 & FP & 14.45 & 0.41 & 42.59 \\
TOI 4191 & 49531 & 742.86 & PC & 19.91 & 1.06 & 83.73 \\
TOI 430 & 18761 & 0.59 & FP & 23.17 & 1.61 & 66.09 \\
TOI 4302 & 4599 & 38.76 & PC & 82.56 & 2.61 & 132.22 \\
TOI 4303 & 22414 & 8.61 & PC & 42.0 & 0.96 & 253.4 \\
TOI 4304 & 41378 & 15.57 & KP & 30.64 & 1.22 & 105.98 \\
TOI 4304 & 41378 & 31.72 & KP & 30.64 & 1.22 & 105.98 \\
TOI 4305 & 102133 & 183.0 & PC & 14.12 & 0.51 & 161.04 \\
TOI 4305 & 102133 & 374.36 & PC & 14.12 & 0.51 & 161.04 \\
TOI 4307 & 25351 & 32.7 & PC & 1.48 & 0.31 & 36.12 \\
TOI 4307 & 25351 & 4.65 & PC & 1.48 & 0.31 & 36.12 \\
TOI 4309 & 51743 & 87.22 & PC & 37.57 & 2.23 & 76.06 \\
TOI 431 & 26013 & 12.46 & CP & 4.22 & 0.85 & 32.62 \\
TOI 431 & 26013 & 0.49 & CP & 4.22 & 0.85 & 32.62 \\
TOI 4320 & 16038 & 703.62 & PC & 11.45 & 0.85 & 79.43 \\
TOI 4320 & 16038 & 46.41 & FA & 11.45 & 0.85 & 79.43 \\
TOI 4321 & 107911 & nan & PC & 15.13 & 0.84 & 109.32 \\
TOI 4324 & 47619 & 6.25 & PC & 10.41 & 1.43 & 17.06 \\
TOI 4326 & 115828 & nan & PC & 15.94 & 1.3 & 57.77 \\
TOI 4328 & 21223 & 703.79 & PC & 3.9 & 0.83 & 25.02 \\
TOI 4330 & 71815 & 3.35 & PC & 138.19 & 1.76 & 287.19 \\
TOI 4337 & 53534 & 2.29 & PC & 8.32 & 0.54 & 64.83 \\
TOI 4350 & 10229 & 4.88 & PC & 4.93 & 0.26 & 103.38 \\
TOI 4355 & 31179 & 674.23 & PC & 19.65 & 1.62 & 76.36 \\
TOI 4358 & 113293 & 390.46 & PC & 27.66 & 2.23 & 66.28 \\
TOI 4362 & 34209 & 7.55 & PC & 19.13 & 0.92 & 134.6 \\
TOI 4369 & 32099 & 13.58 & PC & 197.03 & 0.25 & 283.99 \\
TOI 4382 & 86844 & 10.69 & PC & 145.03 & 2.79 & 166.34 \\
TOI 440 & 25670 & 1.08 & FP & 12.13 & 1.33 & 49.37 \\
TOI 444 & 19950 & 17.96 & PC & 5.34 & 0.41 & 57.45 \\
TOI 4470 & 98505 & 2.22 & KP & 6.37 & 2.58 & 19.78 \\
TOI 4481 & 102401 & 0.93 & PC & 4.92 & 1.5 & 12.06 \\
TOI 4498 & 96902 & 5.31 & FP & 42.77 & 2.8 & 79.83 \\
TOI 4517 & 115752 & 1.21 & KP & 25.89 & 2.84 & 29.65 \\
TOI 4524 & 15249 & 0.93 & CP & 14.19 & 1.06 & 63.86 \\
TOI 4527 & 6069 & 0.4 & PC & 12.56 & 0.48 & 18.1 \\
TOI 4537 & 112100 & 6.66 & PC & 35.06 & 2.66 & 70.61 \\
TOI 454 & 15407 & 18.08 & APC & 19.11 & 1.25 & 79.03 \\
TOI 4580 & 79180 & 0.92 & PC & 2.34 & 0.24 & 67.65 \\
TOI 4588 & 92247 & 13.12 & PC & 14.07 & 0.35 & 226.82 \\
TOI 4597 & 22838 & 4.67 & PC & 66.83 & 2.82 & 124.2 \\
TOI 4599 & 31635 & 2.77 & CP & 1.28 & 0.49 & 10.0 \\
TOI 4599 & 31635 & 5.71 & CP & 1.28 & 0.49 & 10.0 \\
TOI 4602 & 18841 & 3.98 & PC & 12.48 & 1.18 & 62.81 \\
TOI 461 & 11865 & 10.92 & PC & 14.14 & 1.22 & 45.74 \\
TOI 4612 & 29301 & 4.11 & KP & 10.87 & 0.46 & 134.54 \\
TOI 4626 & 66854 & 17.48 & PC & 12.13 & 1.21 & 51.15 \\
TOI 4631 & 45012 & 33.64 & PC & 15.48 & 1.52 & 62.26 \\
TOI 4641 & 13224 & 22.1 & PC & 31.72 & 1.87 & 87.65 \\
TOI 469 & 29442 & 13.63 & CP & 24.04 & 2.03 & 68.0 \\
TOI 480 & 27849 & 6.87 & PC & 22.21 & 2.84 & 54.28 \\
TOI 486 & 31300 & 1.74 & PC & 4.28 & 0.76 & 15.22 \\
TOI 500 & 34269 & 0.55 & CP & 7.15 & 0.71 & 47.41 \\
TOI 5076 & 15683 & 23.44 & PC & 33.03 & 1.05 & 82.75 \\
TOI 5082 & 34271 & 4.24 & PC & 10.69 & 0.98 & 43.03 \\
TOI 509 & 38374 & 9.06 & CP & 13.6 & 1.23 & 48.85 \\
TOI 509 & 38374 & 21.4 & CP & 13.6 & 1.23 & 48.85 \\
TOI 5099 & 13913 & 14.45 & PC & 70.32 & 2.65 & 91.99 \\
TOI 5108 & 54186 & 6.75 & PC & 67.85 & 1.48 & 130.97 \\
TOI 5125 & 27695 & 5.91 & PC & 24.13 & 0.57 & 165.4 \\
TOI 5128 & 46885 & 7.6 & PC & 70.07 & 1.08 & 192.93 \\
TOI 5141 & 50496 & 11.81 & KP & 48.13 & 1.68 & 131.36 \\
TOI 5156 & 17668 & 22.85 & PC & 178.84 & 1.75 & 159.46 \\
TOI 5384 & 51260 & 2.99 & KP & 20.02 & 0.26 & 218.06 \\
TOI 5387 & 69858 & 2.8 & PC & 22.15 & 1.03 & 141.19 \\
TOI 5392 & 69240 & 17.53 & PC & 21.85 & 2.22 & 48.51 \\
TOI 5394 & 50469 & 15.19 & PC & 43.9 & 2.88 & 64.18 \\
TOI 5401 & 61637 & 6.83 & PC & 137.4 & 1.4 & 216.55 \\
TOI 554 & 18893 & 7.05 & PC & 2.83 & 0.31 & 45.18 \\
TOI 554 & 18893 & 3.04 & PC & 2.83 & 0.31 & 45.18 \\
TOI 560 & 42401 & 6.4 & CP & 10.09 & 1.36 & 31.59 \\
TOI 560 & 42401 & 18.88 & CP & 10.09 & 1.36 & 31.59 \\
TOI 562 & 47103 & 3.93 & CP & 4.34 & 1.86 & 9.44 \\
TOI 5724 & 80264 & 697.4 & PC & 1.9 & 0.26 & 50.4 \\
TOI 5739 & 66730 & 8.43 & PC & 19.63 & 1.87 & 61.84 \\
TOI 5789 & 99452 & 12.93 & PC & 2.52 & 0.9 & 20.44 \\
TOI 5807 & 101511 & 14.24 & PC & 10.79 & 1.12 & 73.06 \\
TOI 5809 & 103502 & 9.21 & PC & 229.51 & 1.73 & 38.67 \\
TOI 5817 & 106097 & 15.61 & PC & 16.37 & 1.09 & 80.34 \\
TOI 5821 & 104513 & 2.15 & KP & 98.31 & 2.73 & 182.19 \\
TOI 585 & 43991 & 5.55 & APC & 24.1 & 0.94 & 157.55 \\
TOI 587 & 42654 & 8.04 & PC & 37.57 & 0.96 & 210.31 \\
TOI 588 & 33609 & 39.47 & PC & 63.0 & 1.98 & 151.88 \\
TOI 5951 & 113300 & 3.17 & PC & 65.87 & 2.72 & 160.83 \\
TOI 5955 & 116907 & 0.59 & PC & 58.48 & 1.95 & 43.07 \\
TOI 5961 & 114941 & 1.62 & PC & 9.09 & 1.18 & 26.29 \\
TOI 5972 & 108859 & 3.52 & KP & 7.35 & 0.7 & 48.15 \\
TOI 5997 & 85850 & 5.66 & PC & 21.2 & 2.07 & 46.62 \\
TOI 6026 & 74 & 1.28 & PC & 8.73 & 1.2 & 42.0 \\
TOI 6054 & 17540 & 7.49 & PC & 24.03 & 1.64 & 78.73 \\
TOI 6054 & 17540 & 12.58 & PC & 24.03 & 1.64 & 78.73 \\
TOI 6075 & 91906 & 832.92 & PC & 7.89 & 0.84 & 39.88 \\
TOI 6098 & 38729 & 2.73 & PC & 27.15 & 2.18 & 81.72 \\
TOI 651 & 29118 & 1.07 & PC & 30.07 & 2.03 & 85.86 \\
TOI 652 & 48739 & 3.98 & CP & 9.51 & 1.02 & 45.6 \\
TOI 653 & 47371 & 0.69 & FP & 94.5 & 1.99 & 213.86 \\
TOI 664 & 52733 & 4.74 & KP & 10.75 & 0.48 & 99.95 \\
TOI 704 & 28754 & 3.81 & CP & 2.78 & 0.21 & 29.8 \\
TOI 731 & 47780 & 0.32 & CP & 19.11 & 1.64 & 9.42 \\
TOI 740 & 49678 & 2.13 & PC & 21.34 & 0.87 & 114.55 \\
TOI 741 & 45908 & 7.58 & PC & 0.46 & 0.23 & 10.45 \\
TOI 755 & 61820 & 2.54 & CP & 38.19 & 1.64 & 105.82 \\
TOI 801 & 32674 & 0.78 & PC & 21.42 & 1.73 & 71.4 \\
TOI 802 & 31134 & 3.69 & PC & 1.31 & 0.26 & 28.06 \\
TOI 836 & 73427 & 8.6 & CP & 9.38 & 1.19 & 27.51 \\
TOI 836 & 73427 & 3.82 & PC & 9.38 & 1.19 & 27.51 \\
TOI 869 & 16521 & 26.48 & FA & 26.0 & 1.36 & 110.53 \\
TOI 911 & 85583 & 8.58 & APC & 60.76 & 2.94 & 85.17 \\
TOI 957 & 24689 & 0.83 & FP & 39.77 & 0.64 & 275.31 \\
\hline
\end{longtable*}
\tablenotetext{a}{The flags come from NASA Exoplanet Archive. CP: Confirmed Planet, PC: Planet Candidate, APC: Ambiguous Planet Candidate, FP: False Positive, FA: False Alarm.  }
\end{center}

\newpage


\bibliography{sample63}{}
\bibliographystyle{aasjournal}


\end{CJK*}
\end{document}